\documentclass[10pt,preprint2]{aastex}
\usepackage{amsmath}
\usepackage{graphicx}
\usepackage{color}
\usepackage{mathrsfs}
\usepackage{geometry}
\usepackage{threeparttable}

\newcommand{\km}{\,{\rm km}} \newcommand{\cm}{\,{\rm cm}}
\newcommand{\erg}{\,{\rm erg}} 
\newcommand{\ps}{\,{\rm s}^{-1}}
\newcommand{\kpc}{\,{\rm kpc}}
\newcommand{\parsec}{\,{\rm pc}}

\newcommand{\pcc}{\,{\rm cm}^{-3}}
\newcommand{\psc}{\,{\rm cm}^{-2}}

\newcommand{\snr}{W50}
\newcommand{\NHH}{N({\rm H}_2)}
\newcommand{\HH}{({\rm H}_2)}
\newcommand{\twCO}{$^{12}$CO}
\newcommand{\thCO}{$^{13}$CO}

\newcommand{\otz}{$J$=\,1--0}
\newcommand{\tto}{$J$=\,2--1}
\newcommand{\cnj}{$J$=\,3/2--1/2 $F$=\,5/2--3/2}
\newcommand{\VLSR}{V_{\rm LSR}}
\newcommand{\du}{d_{3.5}}

\newcommand{\E}[1]{\times 10^{#1}}

\begin{document}
\begin{tiny}

\end{tiny}
\title{A Small-scale Investigation of Molecular Emission toward the
Tip of the Western Lobe of \snr/SS\,433}
\author{Qian-Cheng Liu$^1$; Yang Chen$^{1,2}$; 
Ping Zhou$^{3,1}$; Xiao Zhang$^{1,2}$; Bing Jiang$^{1}$}
\affil{$^{1}$Department of Astronomy, Nanjing University, 
163 Xianlin Avenue, Nanjing 210023, China; ygchen@nju.edu.cn}
\affil{$^{2}$Key Laboratory of Modern Astronomy and Astrophysics,
Nanjing University, Ministry of Education, Nanjing 210023, China}
\affil{$^{3}$Anton Pannekoek Institute, University of Amsterdam,
P.O. Box 94249, 1090 GE Amsterdam, The Netherlands}

\begin{abstract}
We perform a molecular (CO and CN) line observation using 
the IRAM\,30m telescope
toward two small regions near the western edge of supernova remnant
(SNR) \snr/SS\,433.
CO observation reveals spatial correspondence of two molecular clumps
at the local-standard-of-rest (LSR) velocity around $+53\km\ps$,
with multiwavelength local features of the W50/SS\,433 system.
One of the two clumps appears to be embedded in a void of diffuse
radio and X-ray emission.
Toward the two clumps, asymmetric broad-line profiles of the
\twCO\ lines are obtained, which provide kinematic evidence
of the association between the clumps and the jet-related gas.
The \twCO\ \tto/\otz\ line ratios ($\ga 0.9$) and the kinetic
temperatures ($\sim 30\,{\rm K}$) of the
clumps are distinctively higher than all those of the clumps at
other LSR velocities along the same line of sight, 
which may be physical signatures of the association.
We show that the clump coincident with the void can survive the thermal heating 
if it is surrounded by hot gas, with an evaporation timescale much 
larger than the age of SNR~\snr. 
We also show that the thermal equilibrium in the high-temperature 
clumps can be maintained by the heating of the penetrating environmental 
cosmic rays.
CN ($J=3/1$--$1/2$) line emission is detected in the two clumps, 
and the CN abundances derived 
are much higher than that in the interstellar molecular clouds (MCs)
and that in the SNR-interacting MCs.
\end{abstract}

\keywords{Supernova remnants (1667); Interstellar medium (847); 
Molecular clouds (1072); Jets (870)}

\section{Introduction} 

Microquasar SS\,433 is the first discovered binary system containing a 
stellar-mass compact object with relativistic jets 
\citep{1979Natur.279..701A,1979MNRAS.187P..13F}.
It is located in the center of radio shell \snr,
a manatee-like supernova remnant (SNR) cataloged as
SNR G39.7-2.0 \citep{2014BASI...42...47G},
of which the radio size is $\sim 120'\times 60'$.
The \snr/SS\,433 system is suggested to interact with the interstellar medium 
\citep[ISM; e.g.,][]{1983ApJ...272..609H,1990A&A...240...98W,
1998AJ....116.1842D,2000AdSpR..25..703D,2007MNRAS.381..881L,
2008PASJ...60..715Y,2018ApJ...863..103S}
and induce molecular clouds (MC) formation 
\citep[e.g.,][]{2008PASJ...60..715Y,
2014ApJ...789...79A,2018ApJ...863..103S}.
Except for the \snr/SS\,433 system, there are very few observational 
examples in our Galaxy of the possible association between 
relativistic jets and MCs
\citep[e.g.,][]{2014ApJ...781...70F}.
On a larger scale, active galactic nucleus radio jet feedback could
have an important impact on the ISM of the host galaxy
and affect the process of star formation.
However, it is unclear how molecular gas is affected
by the jet. 
Hence, the study of the microquasar in our Galaxy, like the
SS\,433/\snr\ system, toward which possible association with 
MC has been suggested, could provide an opportunity for a detailed, 
deep case study of this kind of physical process.

The jet on each side of SS\,433 is extended by $\sim 1^\circ$
\citep{1998PhDT..........K}, an angular size that can be 
translated to a length of about $50$--$80\parsec$
at an estimated distance of $\sim 3$--$5\kpc$
\citep[e.g.,][]{1981ApJ...246L.141H,
1998AJ....116.1842D,2004ApJ...616L.159B,2008PASJ...60..715Y,
2013ApJ...775...75M,2014A&A...562A.130P,2018ApJS..238...35S,
2018ApJ...863..103S}.
In the radio band, the large-scale environment of \snr\ has been
studied \citep[e.g.,][]{1980A&A....84..237G,1981A&A....97..296D,
1986MNRAS.218..393D,1998AJ....116.1842D,2011A&A...529A.159G},
and the radio ears of \snr\ were suggested to be the result 
of the interaction between the ram pressure of the jets 
and the SNR shell \citep[e.g.,][]{1987AJ.....94.1633E}.
In far-infrared (IR), six knots were found
\citep{1987PASP...99.1269B,1989ApJ...338..945B}
and suggested to be
a result of jet-clump interaction 
\citep{1990A&A...240...98W}.
In X-rays, 
two bright, diffuse lobes were found \citep{1983ApJ...273..688W}, and
a possible interaction between the jets and their
ambient gas was studied using data from 
ROSAT, ASCA, RXTE, and Chandra 
\citep{1994PASJ...46L.109Y,1996A&A...312..306B,1997ApJ...483..868S, 
1999ApJ...512..784S, 2005AdSpR..35.1062M,2007A&A...463..611B}.
In gamma-rays, emissions have been detected by the
Fermi Large Area Telescope \citep[LAT; e.g.,][]{2015ApJ...807L...8B}
toward SS\,433, by the High Altitude Water Cherenkov observatory 
\citep{2018Natur.562...82A} toward the lobes of SS\,433,
and no evidence of gamma-ray emission from the jet termination regions
has been found between a few hundred GeV and a few TeV 
\citep{2018A&A...612A..14M}. 

\snr\ was suggested to lie at the end of a filamentary MC at 
local-standard-of-rest (LSR) velocity $\VLSR \sim +32\km\ps$
and be related to the MC \citep{1983ApJ...272..609H}.
A scenario of SNR-HI interaction at $\VLSR \sim +42\km\ps$ was
suggested according to the observation of radio continuum emission at
1.4\,GHz and emission of HI 21\,cm \citep{1998AJ....116.1842D}.
An MC at 
$\VLSR \sim +50\km\ps$
was found to be spatially coincident with the X-ray hot spots
toward one part of the jets 
\citep{2000AdSpR..25..703D}.
The shapes of two MCs at $\VLSR \sim +50\km\ps$ were found to be similar
to the $15\mu m$ mid-IR emission shape \citep{2002astro.ph..7429F}.
Furthermore, 10 MCs were found at $\VLSR=+40$ -- $+60\km\ps$, 
which align nearly along the axis of the jets of SS\,433 
based on an analysis of the NANTEN \twCO\ (\otz) data
\citep{2008PASJ...60..715Y}.
Accordingly, it was proposed that they are created by the interaction
of relativistic jets and ambient interstellar HI gas.
However, the low spatial resolution ($2'.6$) and grid spacing ($4'$)
of the data used in the paper prevented the study of the kinematic/physical
features of the clumps.
In addition, H$^{13}$CO$^+$
molecules were found at the western part
of the \snr/SS\,433 system at $\VLSR=+50$ -- $+55\km\ps$
\citep{2001A&A...366.1035C}.
Nonetheless, it is also argued that the system is associated with
the gas material at an LSR velocity of $\sim +75\km\ps$ according to
observations of HI, CO and optical emission 
\citep{2007MNRAS.381..308B,2007MNRAS.381..881L, 2018ApJ...863..103S}.

The interaction between jets and the ISM has also been studied by 
hydrodynamic and magnetohydrodynamic (MHD) simulations,
and it is shown that MCs can be formed 
\citep[e.g.,][]{2014ApJ...789...79A,2017ApJ...836..213A}
with either an arc-like shape or a line-like shape that aligns
with the jet, depending on the filling factor of the HI clumps
that the jet interacts with \citep{2017ApJ...836..213A}.
Although the assumed velocities of the jets in the simulations
(about hundreds of $\km\ps$) are much smaller than that for SS 433
\citep[$\sim 0.26\,c$;][]{1989ApJ...347..448M}, these results have
demonstrated that the jets of SS\,433 do have the potential
to induce MC formation, especially when penetrating
into a dense HI environment (e.g., the western 
lobe of \snr,
which is closer to the disk of the Milky Way).

Although there are some observational results that are
consistent with the scenario of 
interaction between molecular gas and the
\snr/SS\,433 system 
\citep[e.g.,][]{2008PASJ...60..715Y,2018ApJ...863..103S},
it is still an open question whether there is association and 
at which LSR velocity the association could be 
\citep{1989ApJ...338..945B,1998AJ....116.1842D, 2000AdSpR..25..703D,
2001A&A...366.1035C, 2005AdSpR..35.1062M, 2007MNRAS.381..308B,
2007MNRAS.381..881L, 2008PASJ...60..715Y, 2018ApJ...863..103S}.
There is some observational evidence for judging the interaction
with MCs, such as molecular line broadening, 
high high-to-low excitation line ratio,
and morphological agreement
\citep{2010ApJ...712.1147J, 2014IAUS..296..170C}.
Motivated by the observational study of the 10 molecular clumps in
\cite{2008PASJ...60..715Y} and simulational studies of
distribution of the molecular gas interacting with a jet
\citep[e.g.,][]{2014ApJ...789...79A,2017ApJ...836..213A},
we have performed an IRAM\,30m observation toward 2 of 
the 10 clumps (SS\,433-N3 and part of SS\,433-N2 in
\citealt{2008PASJ...60..715Y}) in molecular lines.
The two clumps are located at the 
western tip of the X-ray 
and radio emission, which are suggested to result from
the interaction of the jet of SS\,433 and the surrounding 
medium \citep[e.g.,][]{1997ApJ...483..868S,1998AJ....116.1842D}.
The two clumps were selected based on the following 
two considerations.
(1) They are located at low Galactic latitude,
where the HI density should be higher 
than at the other side, making jet-induced 
CO formation more plausible \citep[e.g.,][]{2017ApJ...836..213A}.
(2) Since they are located at the tip of the radio emission,
any possible morphological correspondence could more easily be observed.
In this paper, we report CO and CN line observation
toward two clumps, aiming to look for the signature 
of the interaction of the \snr/SS\,433 system with the MCs
in a higher spatial resolution. We describe the 
observations of the CO line emissions and the data reduction
process in \S 2; we present the results of our CO data
in detail in \S3; and discuss the main results in \S4.
We summarize this paper in \S5.

\begin{table}[t]
\centering
\footnotesize
\begin{threeparttable}
\caption{\footnotesize Observational information of regions N2 and
N3\tnote{a} \label{targets}}
\begin{tabular}{cccc}
\hline \\
 & & & Velocity \\
Region & Line &
Frequency & Resolution \\
 & & (GHz) & ($\km\ps$) \\
\hline\\
N2   & \twCO (\otz) & 115.271 & 0.508 \\
N2   & \thCO (\otz) & 110.201 & 0.508 \\
N2   & \twCO (\tto) & 230.538 & 0.254 \\
N2   & CN ($J=3/2-1/2$ & 113.491 & 0.508 \\
     & $F=5/2-3/2$) \\
N3   & \twCO (\otz) & 115.271 & 0.127/0.508 \\
N3   & \thCO (\otz) & 110.201 & 0.508 \\
N3   & \twCO (\tto) & 230.538 & 0.063/0.254 \\
N3   & CN ($J=3/2-1/2$ & 113.491 & 0.508 \\
     & $F=5/2-3/2$) \\
\hline
\end{tabular}
\begin{tablenotes}
\item[a] \footnotesize The regions are defined in \S\ref{data}
and delineated in Figure~\ref{overallspec}.
\end{tablenotes}
\end{threeparttable}
\end{table}

\begin{figure*}[t]
\centering
\includegraphics[scale=.35]{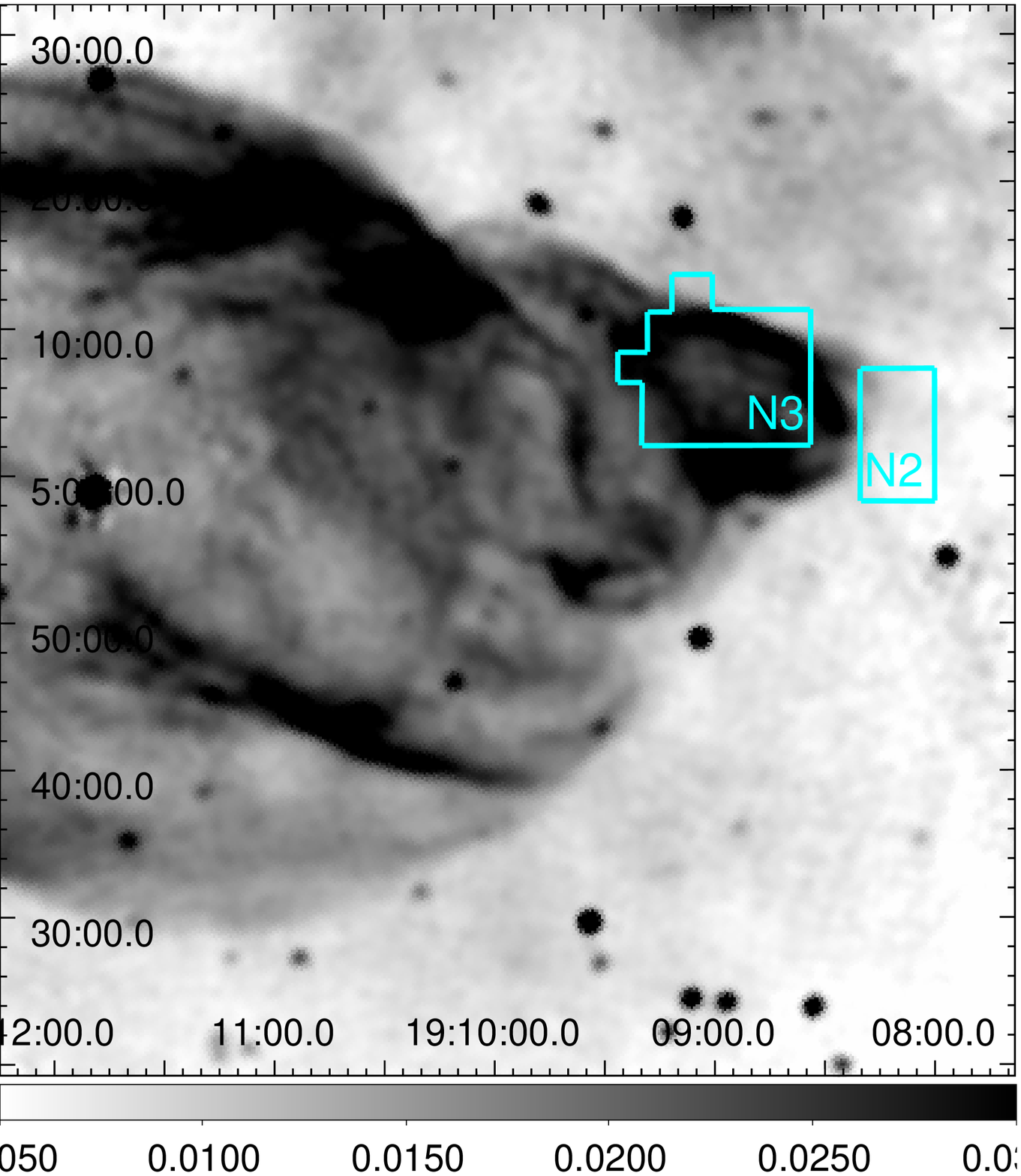}
\includegraphics[scale=.55]{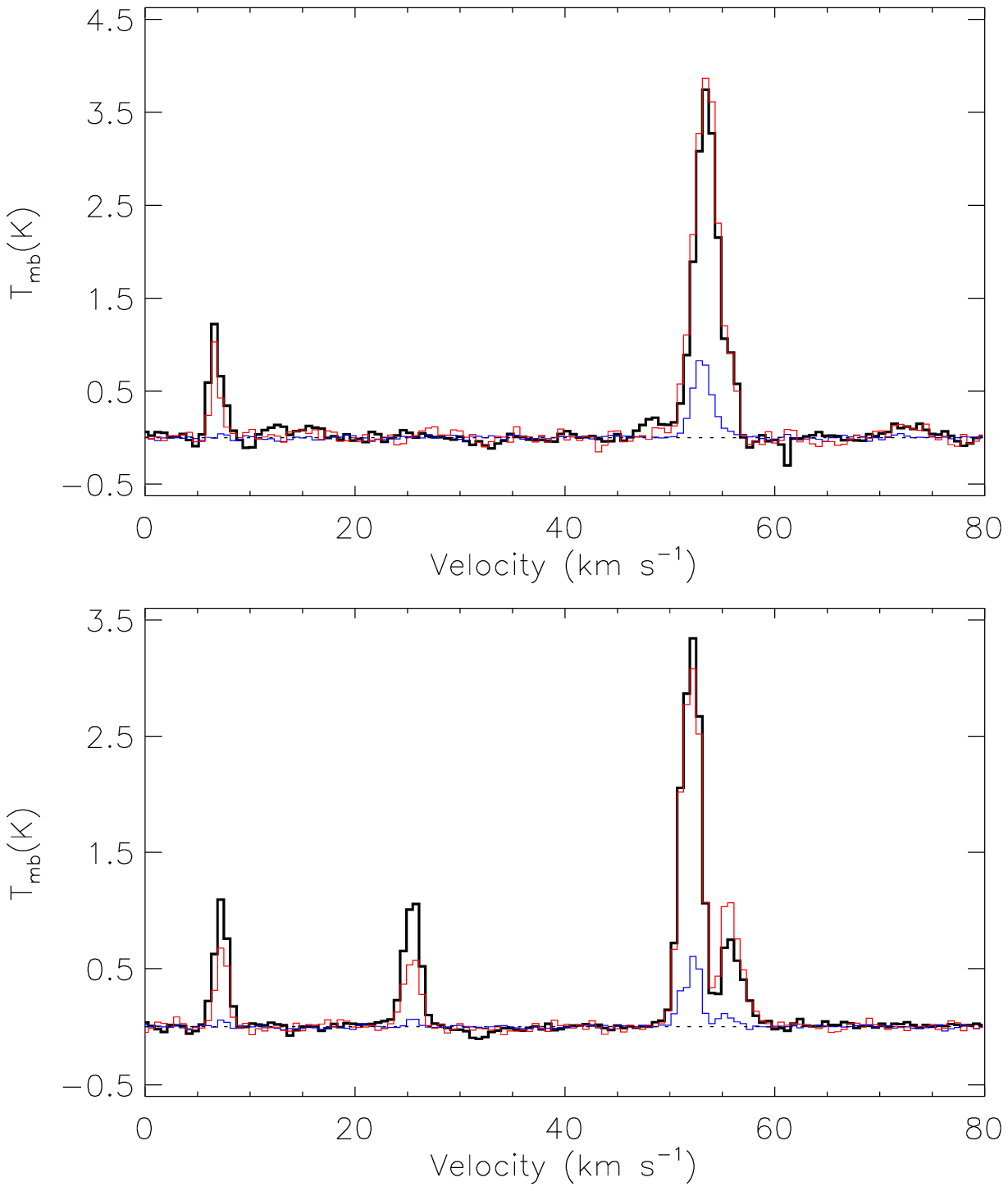}
\caption{\footnotesize
(Left) Western part of SNR\,\snr\ in radio continuum at 1.4\,GHz
from \cite{1998AJ....116.1842D}. Regions labeled with ``N2" and
``N3" delineate the regions we observed with IRAM\,30m.
(Right)
Average CO spectra of regions N2 (right top panel) and 
N3 (right bottom panel) in the velocity range 0 to $+80\km\ps$
from the IRAM observation.
The black thick lines are for \twCO\ (\otz),
the blue lines are for \thCO\ (\otz), the red lines are for \twCO (\tto),
and the dotted lines are for the 0\,K main-beam temperature.
Note that the \twCO\ (\tto) is resampled to achieve the same
velocity resolution $0.6\km\ps$ as the \twCO\ (\otz).
}
\label{overallspec}
\end{figure*}

\section{Observations and data reduction} \label{data}

Our observations of molecular lines toward SNR~\snr\ were made
simultaneously in \twCO\ (\otz), \thCO\ (\otz), and \twCO\ (\tto)
with the IRAM 30\,m telescope
during 2017 December 9--10, with a total of 12 hr.
The observations covered two regions of areas 
$\sim 5'\times 9'$ and $\sim 12'\times 10'$ centered at
($19^{\rm h} 08^{\rm m} 10^{\rm s}, +05^{\circ} 02' 45''$, J2000)
and 
($19^{\rm h} 08^{\rm m} 58^{\rm s}, +05^{\circ} 07' 14''$, J2000)
(regions N2 and N3 hereafter, corresponding to part of 
SS\,433-N2 and complete SS\,433-N3 in \citealt{2008PASJ...60..715Y}),
respectively.
The mappings were conducted with the on-the-fly position switching
mode using the Eight Mixer Receiver in E0 and E1 bands
and the fast Fourier transform spectrometers (FTS).
The backend FTS provided either a bandwidth of 4\,GHz and a 
spectral resolution of 50\,kHz in part of the observation during
2017 December 9 or a bandwidth of 16\,GHz and a spectral resolution
of 200\,kHz. Notably, the emission of CN (\cnj) 
line at 113.491\,GHz, which was recognized in our analysis
(see \S\ref{result}), was covered in the wider 
bandwidth of the 16\,GHz mode.
The observational information, including the 
velocity resolution for each line, is summarized in 
Table~\ref{targets} (note that the \thCO\ (\otz) line was not
covered when a high-velocity resolution of 
$0.127\km\ps$ was applied to \twCO\ (\otz)).
The half-power beamwidths (HPBWs) of the telescope were $\sim 21''$
at 115\,GHz and $\sim 10.7''$ at 230\,GHz, with main-beam efficiencies
of $\sim 78$\% at 115\,GHz and $\sim 59$\% at 230\,GHz, respectively.
All the data observed with the IRAM 30m telescope were reduced using the 
GILDAS/CLASS package developed by the IRAM observatory\footnote{
http://www.iram.fr/IRAMFR/GILDAS}. 
To reduce the noise level, the \twCO\ (\tto) line data were resampled to 
achieve a velocity resolution $0.3\km\ps$, while the data of other lines
(\twCO\ (\otz), \thCO\ (\otz), and CN (\cnj)) were resampled 
to achieve a velocity resolution $0.6\km\ps$. 
Also, the angular resolutions of all line data were
convolved to $22''$.
The mean rms noise levels of the main-beam temperature ($T_{\rm mb}$)
of the data
are $\sim 0.3$/0.2\,K (\twCO\ \otz),
$\sim 0.15$/0.1\,K (\thCO\ \otz, CN \cnj), 
and $\sim 0.23$/0.19\,K
(\twCO\ \tto) for the N2/N3 regions, respectively. 

In addition, we used \twCO\ (\otz)
data from the Milky Way Imaging Scroll Painting (MWISP) project,
which was observed with the 13.7\,m millimeter-wavelength telescope
of the Purple Mountain Observatory at Delingha (PMOD), to show the larger
environment of the N2 and N3 regions.
The HPBW of the data was $\sim 50''$, with velocity resolution
of $\sim 0.16\km\ps$.
The typical mean rms noise level was 0.5\,K.
For the purposes of a multiwavelength investigation of the environment,
we also used ROSAT PSPC X-ray (ObsID: US400271P-1.N1, 
PI: Marshall Dr., Francis E.), Wide-field Infrared Survey Explorer
(WISE) 12/22 $\mu$m mid-IR (WISE Science Data Center, IPAC, Caltech), 
and Very Large Array (VLA) 1.4\,GHz radio continuum 
\citep[from][]{1998AJ....116.1842D} data.

\begin{figure*}
\centering
\includegraphics[scale=.7]{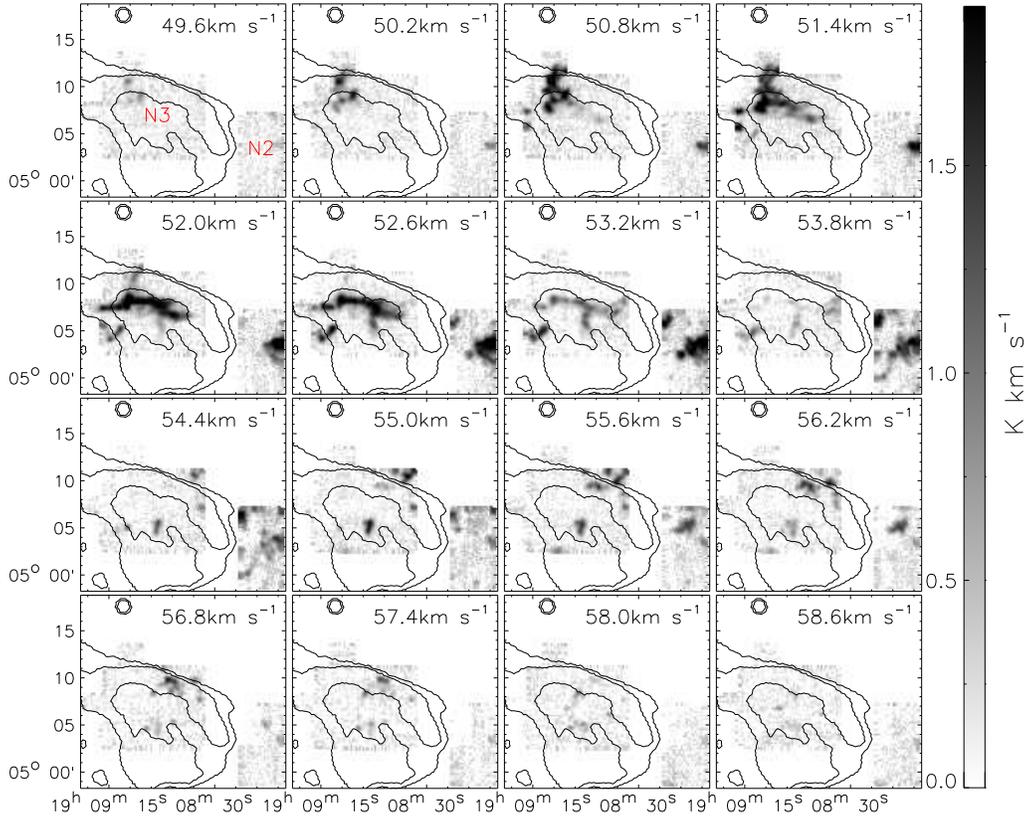}
\caption{\footnotesize IRAM
\twCO\ (\otz) intensity maps integrated at $0.6 \km \ps$
in the velocity range $+49.3$ to $+58.9 \km \ps$,
overlaid by VLA 1.4\,GHz radio contours from 
\cite{1998AJ....116.1842D} in black contours.
The color bar shows the square root of the value,
and an equatorial coordinate system in J2000 is used.
Regions N2 and N3 are labeled in the upper left panel.
\label{channelmap1}}
\end{figure*}

\begin{figure*}
\centering
\includegraphics[scale=.7]{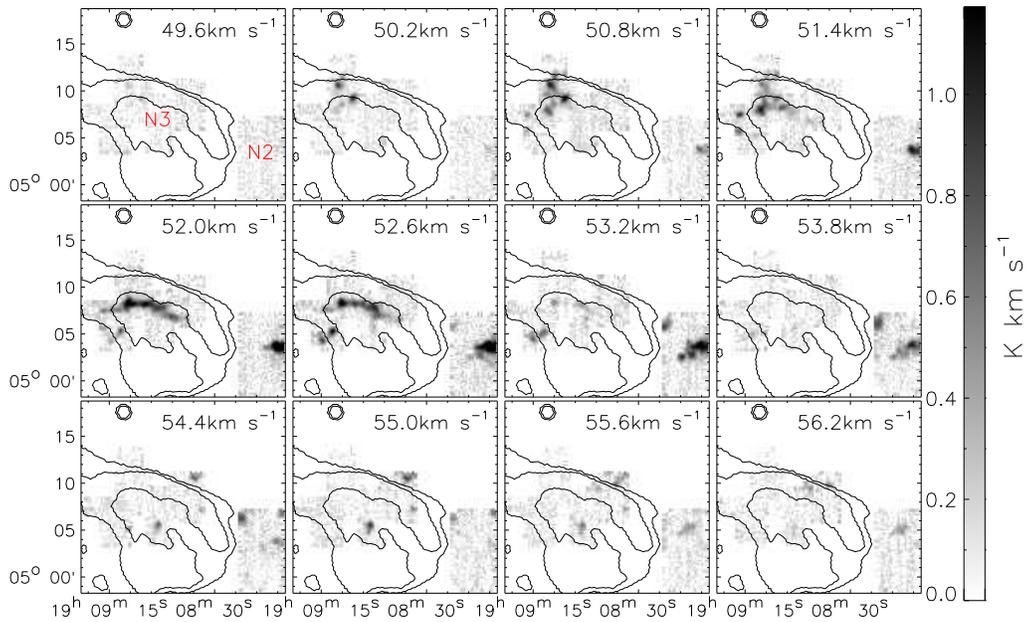}
\caption{\footnotesize IRAM \thCO\ (\otz) intensity maps integrated at $0.6 \km \ps$
in the velocity range $+49.3$ to $+56.5 \km \ps$,
overlaid by VLA 1.4\,GHz radio contours, the same in Figure~\ref{channelmap1},
The color bar shows the square root of the value.
Regions N2 and N3 are labeled in the upper left panel.
\label{channelmap2}}
\end{figure*}

\begin{figure*}[t]
\centering
\includegraphics[scale=.45]{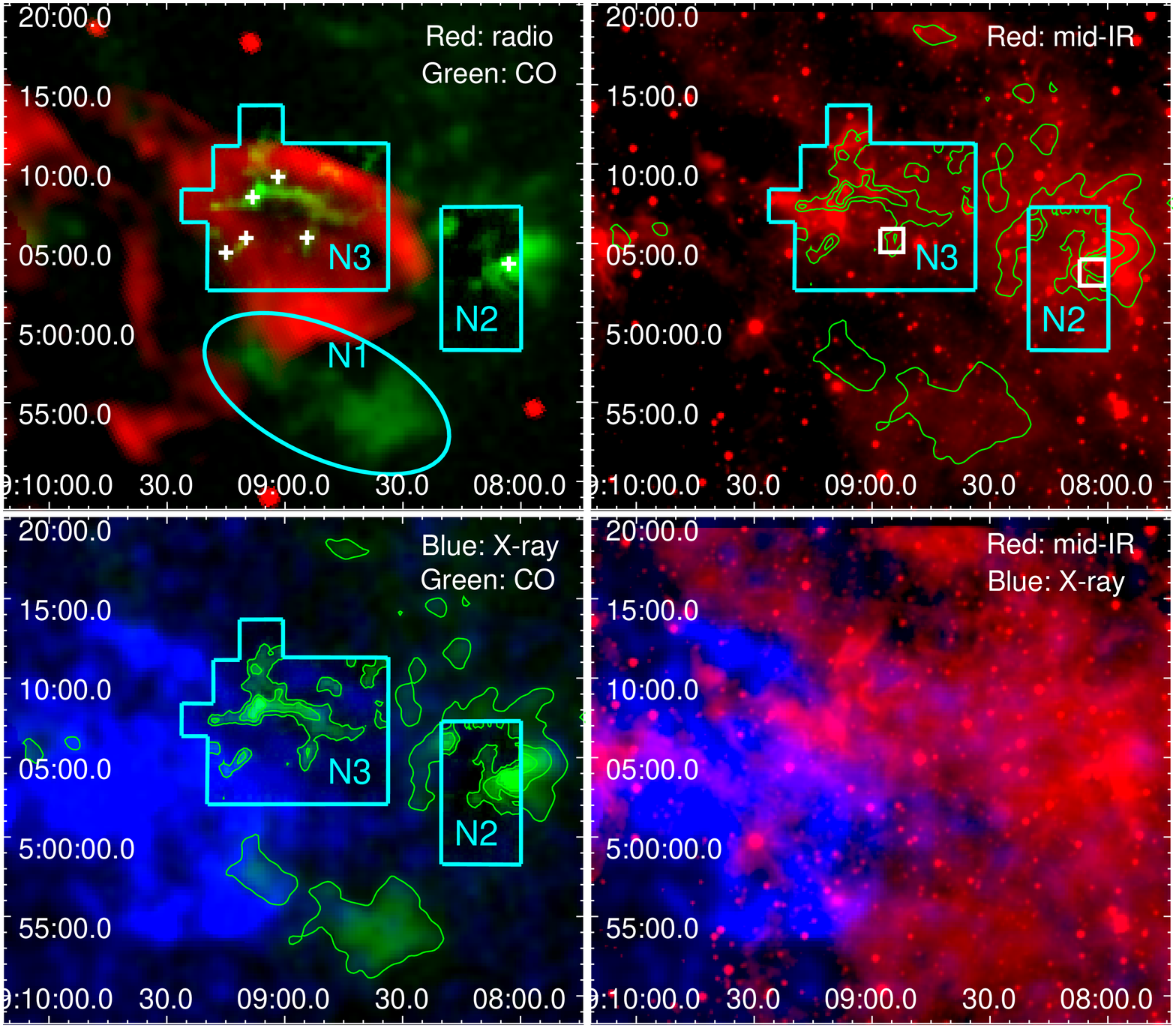}
\caption{\footnotesize Multiwavelength maps of the western lobe of SNR~\snr.
(Upper left) The radio continuum at 1.4\,GHz from 
\cite{1998AJ....116.1842D} is in red, and \twCO\ (\otz) intensity map 
integrated in the velocity range +48 to $+60\km\ps$ is in green 
(the data in regions N2 and N3 are from IRAM\,30\,m, and the data
outside the two regions are from PMOD MWISP).
The ellipse shows the ``SS\,433-N1" clump from
\cite{2008PASJ...60..715Y}.
The six white crosses mark the project positions where CN lines
are detected (see \S~\ref{cnsection}).
(Upper right) The WISE $12.082\mu$m mid-IR image (with a linear scale) is 
in red, overlaid with 
integrated \twCO\ (\otz) in $+48$ to $+60\km\ps$ with levels
10, 20, 30, and 40\,K\,$\km\ps$.
The two regions delineated by white rectangles are used to
extract CO grid spectra (Figure~\ref{linegridn1} and \ref{linegridn2}).
(Bottom left) \twCO\ (\otz) data shown in green are the same as those
in the upper left panel.
ROSAT PSPC 0.5 -- 2\,keV X-ray emission is in blue.
(Bottom right) The WISE $12.082\mu$m mid-IR image
(with a square root scale) is in red,
and ROSAT PSPC 0.5 -- 2\,keV X-ray emission is in blue
(same as in the bottom left panel).
\label{multiwave}}
\end{figure*}

\begin{figure*}[t]
\centering
\includegraphics[scale=1.]{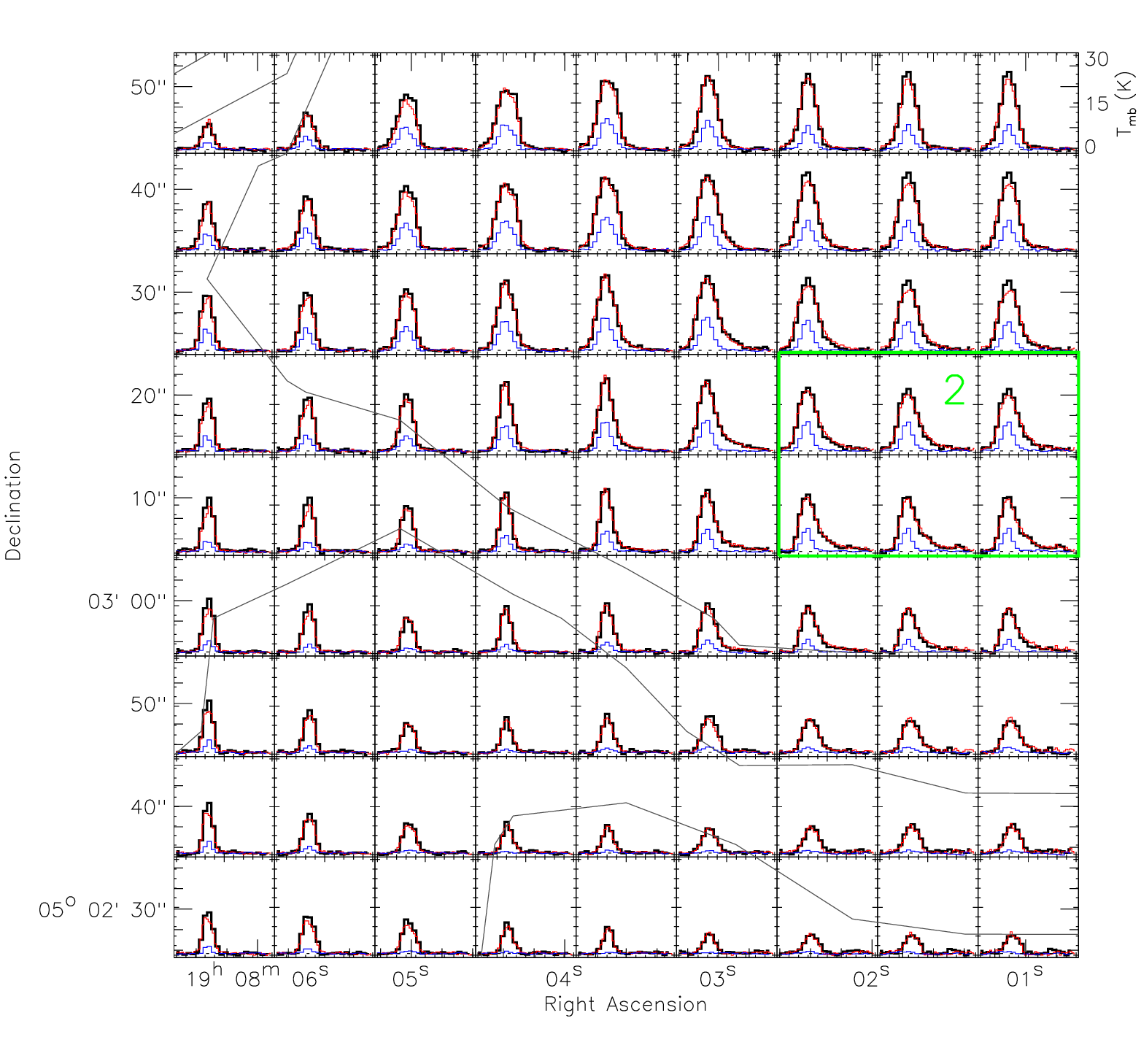}
\caption{\footnotesize Grid of IRAM \twCO\ (\otz), \twCO\ (\tto), and \thCO\ (\otz) 
spectra in the velocity range $+47$ to $+61 \km \ps$
for the region delineated by the white rectangle within 
region N2 in the upper right panel of Figure~\ref{multiwave}.
The black thick lines denote \twCO\ (\otz) spectra, red lines denote
\twCO\ (\tto), blue lines denote \thCO\ (\otz), and dashed lines
denote the 0\,K main-beam temperature.
The size of each pixel is $11''\times 11''$.
The contours are the same as those in Figure~\ref{multiwave}.
\label{linegridn1}}
\end{figure*}

\section{Results} \label{result}

\subsection{Spatial distribution of the clouds}
\label{spatial}
Figure~\ref{overallspec} shows the averaged CO spectra 
from regions N2 
and N3 that were observed with IRAM\,30m.
The CO spectra of region N2 are somewhat different from those of N3.
There are two prominent \twCO~(\otz) and \twCO~(\tto) emission peaks, 
at around $+8$ and $+53 \km\ps$, in region N2,
while the \thCO~(\otz) emission is only prominent at $\sim+53\km\ps$.
On the other hand, four prominent \twCO~(\otz) and \twCO~(\tto)
emission peaks, at around $+8$, $+26$, $+53$,
and $+56 \km\ps$, are shown in region N3.
Only one prominent \thCO~(\otz) emission peak (at $\sim+53\km\ps$) can be seen
in region N3.
Also, the \twCO (\tto)/(\otz) line ratios are higher at 
$\sim +53\km\ps$ than at other velocity ranges in both regions N2 and N3.

We made \twCO\ (\otz) emission channel maps around each velocity 
component to 
examine the spatial distribution of those molecular clumps.
No clear morphological correspondence between
MCs and the SNR was found for CO clumps at
velocities below $+40 \km\ps$, which is consistent with
the results in \cite{2018ApJ...863..103S}.
The \twCO\ (\otz) and \thCO\ (\otz) emission channel maps of the 
$\sim +53\km\ps$ clumps in regions N2 and N3, with a velocity 
interval $0.6\km\ps$, are shown in Figure~\ref{channelmap1} and
Figure~\ref{channelmap2}. 
Some spatial features related to the clumps and the surrounding
multiwavelength emission
are noteworthy. 

(1) Region N2: 
(a) globally, it lies outside of the western tip of
the radio continuum.
(b) In the velocity interval $+53.8$ to $+54.4\km\ps$, there is a weak 
arc-like structure 
in \twCO\ (\otz) emission channel maps (Figure~\ref{channelmap1}). 
The structure, however, does not appear
in the \thCO\ (\otz) channel maps (Figure~\ref{channelmap2}).

(2) Region N3:
(a) there is a remarkable arc-like structure in the velocity interval 
$\sim +51.4$ to $+53.2\km\ps$ (Figure~\ref{channelmap1}) that can also
be discerned in the \thCO\ (\otz) channel maps in velocity interval
$\sim +52$ to $+52.6\km\ps$ (Figure~\ref{channelmap2}).
(b) This clump is anti-correlated with the radio continuum,
appearing embedded in a void of radio emission
(Figure~\ref{channelmap1} and \ref{channelmap2}), which is 
more clearly shown in the upper left panel of Figure~\ref{multiwave}.
The clump is also anticorrelated to the radio continuum at 327.5\,MHz
\citep{1998AJ....116.1842D} and at 150\,MHz
\citep{2018MNRAS.475.5360B}.
It will be shown in \S~\ref{anticor}
that the clump is not likely located in the foreground 
and causing the low radio brightness in region N3 by extinction.
(c) This clump is located on the tip of an X-ray
lobe, and the X-ray emission toward the clump is weaker than 
its surroundings (see the bottom left panel of Figure~\ref{multiwave}).
That is, there also seems to be an anticorrelation between
the clump and the X-ray emission.

(3) Mid-IR view:
although there is a lot of IR emission in the field, which makes
the analysis of the IR emission to be complicated, it is noteworthy
that the
clumps at $\sim +53 \km\ps$ in both regions N2 and N3
are essentially spatially correlated with
two patches of enhanced mid-IR emission (see the upper right panel of
Figure~\ref{multiwave}).
Since the IR emission at $\sim 12\mu m$ seems to be dominated by
aromatic molecules
(e.g., \citealt{2001A&A...372..981V, 2010SCPMA..53S.271W}),
apart from possible contributions from warm dust,
the correlation between the IR emission and the clumps in regions
N2 and N3 may indicate dense gas there.
The mid-IR emission (on a square root scale), on the other hand, 
finely encloses the western edge of the X-ray lobe
(especially the patch along the northwestern edge),
with an anti-correlated brightness distribution with each other
(see bottom right panel of Figure~\ref{multiwave}).
This seems to imply that the dense gas indicated by the
$+53\km\ps$ clumps and the mid-IR emission is associated with
the western X-ray and radio lobe.

\begin{figure*}[t]
\centering
\includegraphics[scale=1.]{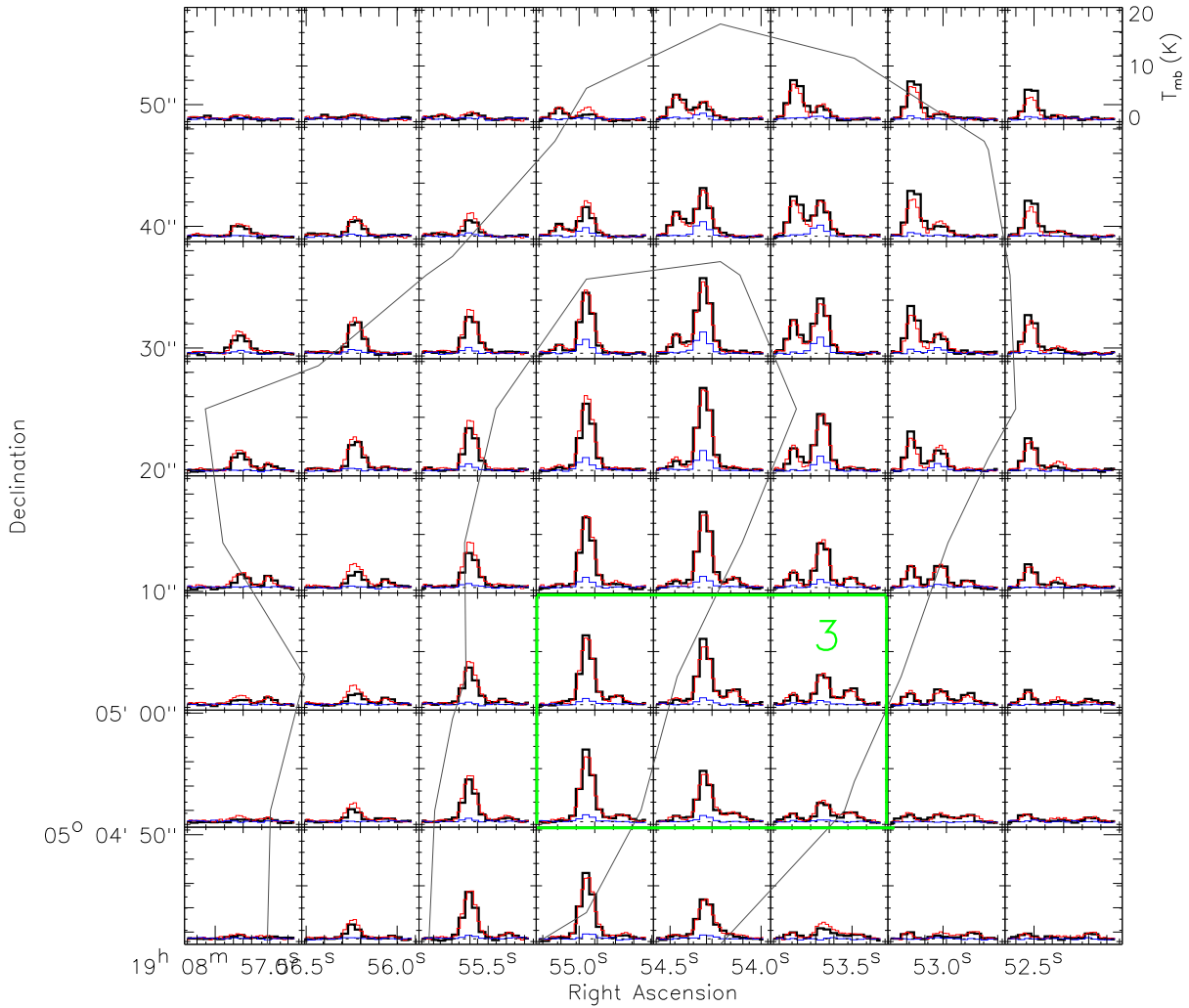}
\caption{\footnotesize Grid of IRAM \twCO\ (\otz), \twCO\ (\tto), and \thCO\ (\otz) 
spectra in the velocity range $+49$ to $+61 \km \ps$
for the region delineated by the white rectangle within
region N3 in the upper right panel of Figure~\ref{multiwave}.
The black thick, red, blue, and dashed lines denote the same spectra as
in Figure~\ref{linegridn1}.
The size of each pixel is $11''\times 11''$.
The contours are also the same as those in Figure~\ref{multiwave}.
\label{linegridn2}}
\end{figure*}

\begin{table*}
\centering
\footnotesize
\begin{threeparttable}
\caption{\footnotesize Observed and Derived Parameters for
the MCs in $+50$ to $+60\km\ps$ in Regions N2 and N3 \tnote{a}
\label{parameter}}
\begin{tabular}{lcccccc}
\hline \\
& & N2~~~~~ & & & N3~~~~~ & \\
& (\twCO\ \otz) & (\twCO\ \tto) & (\thCO\ \otz) 
& (\twCO\ \otz) & (\twCO\ \tto) & (\thCO\ \otz) \\
\hline\\
$T_{\rm peak}$(K) & 25.0 & 22.9 & 10.4 & 28.0 & 26.4 & 9.2 \\
$V_{\rm cent}(\km\ps)$ & 52.7 & 52.9 & 52.8 & 52.1 & 52.1 & 52.5 \\
FWHM ($\km\ps$) & 2.6 & 3.2 & 2.1 & 1.7 & 1.7 & 1.3 \\
$W$(CO) (K\,$\km\ps$) & 70.4 & 77.6 & 23.6 & 49.3 & 47.3 & 12.2 \\
$T_{\rm ex}$, $T_{c,k}$ (K) & 28 & 28 & ... & 31 & 32 & ... \\
$\tau$(\thCO) & ... & ... & 0.56 & ... & ... & 0.34 \\
\hline\\
$N$(H$_2$)($10^{21}$cm$^{-2}$)\tnote{b} & ... & ... & 6.7 
& ... & ... & 3.4 \\
$n(\mbox{H}_2)\du(\cm^{-3})$\tnote{c} & ... & ... & $\sim 1900$ 
& ... & ... & $\sim 900$ \\
$M\du^{-2}(10^2M_\odot)$\tnote{d} & ... & ... & $\sim 4.9$ 
& ... & ... & $\sim 5.0$\\
\hline
\end{tabular}
\begin{tablenotes}
\item[a] \footnotesize The regions are defined in 
\S\ref{data} and delineated in
Figure~\ref{multiwave}.
\item[b] The averaged column densities within each region.
\item[c] The averaged densities within each region.
\item[d] The $\NHH$ masses estimated within each region.
\end{tablenotes}
\end{threeparttable}
\end{table*}

\begin{figure}
\centering
\includegraphics[scale=.5]{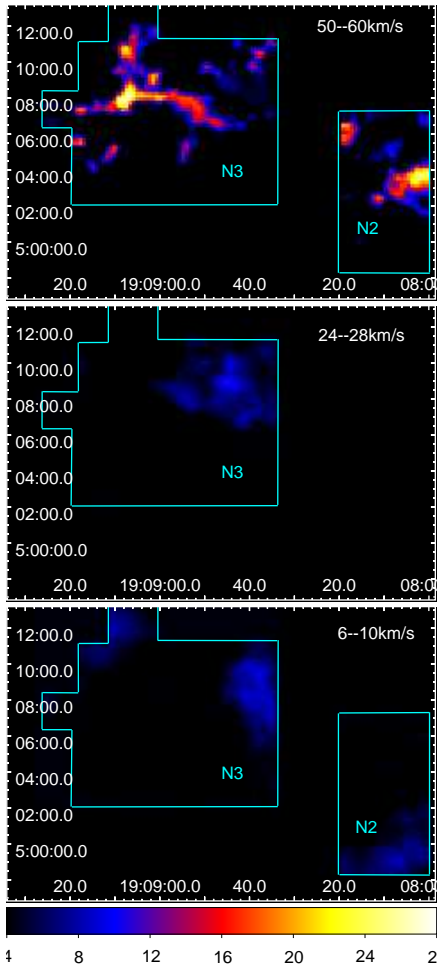}
\caption{\footnotesize Kinetic temperature ($T_{\rm c,k}$) maps of the molecular
clumps at velocities $+50$ to $+60\km\ps$
(top panel), $+24$ to $+28\km\ps$ (middle panel), and $+6$ to $+10\km\ps$ 
(bottom panel) obtained from the IRAM observation,
with assumptions that
there is no beam dilution ($f=1$) and $\tau _{^{12}{\rm CO}}\gg 1$.
Here we have clipped the map where 
the temperature is
lower than 4\,K.
The cyan regions labeled with ``N2" and ``N3" are the same as
in Figure~\ref{overallspec}.
\label{tkin}}
\end{figure}

\begin{figure*}[t]
\centering
\includegraphics[scale=.5]{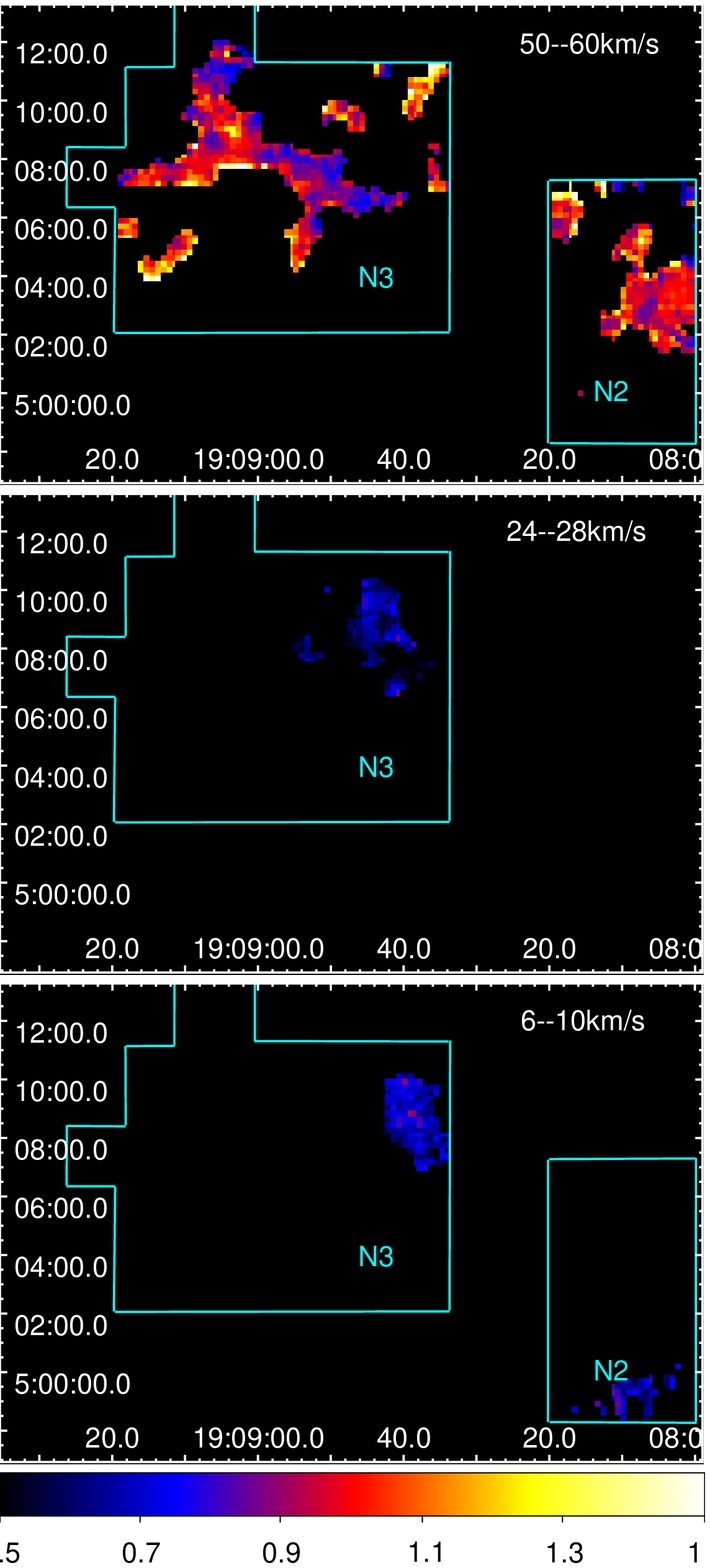}
\includegraphics[scale=.5]{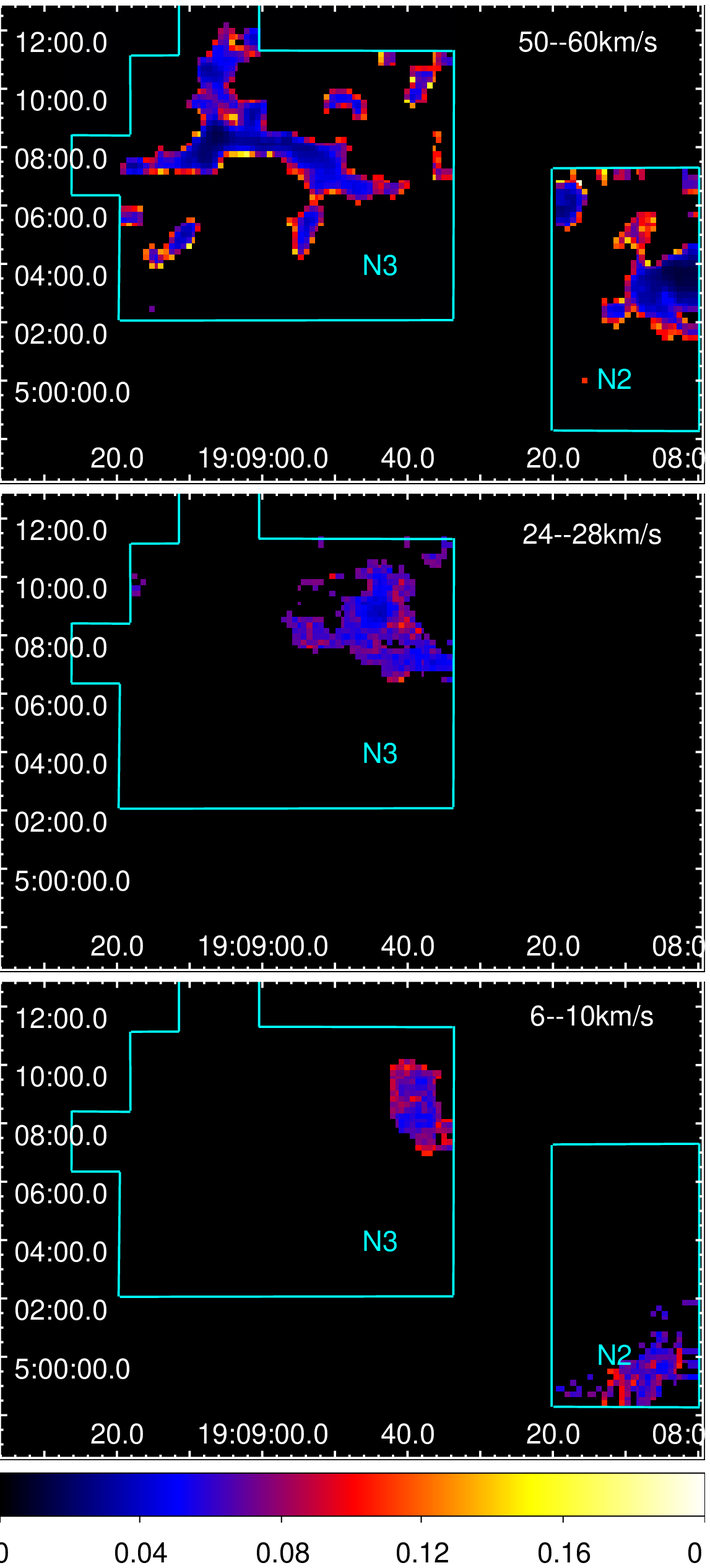}
\caption{\footnotesize
IRAM \twCO\ \tto/\otz\ ratio maps for LSR velocity ranges
$+50$ to $+60\km\ps$ (top left panel), $+24$ to $+28\km\ps$
(middle left panel), and $+6$ to $+10\km\ps$ (bottom left panel),
with the $1\sigma$ error maps shown in the right panels. 
The cyan regions labeled with ``N2" and ``N3"
are the same as those in Figure~\ref{overallspec}.
\label{ratio}}
\end{figure*}

In addition to the two clumps described above,
there is another molecular clump in region N1
south to the tip of the western lobe
(as shown in Figure~\ref{multiwave})
that is identical to the clump named as ``SS\,433-N1" in
\cite{2008PASJ...60..715Y}.
Three clumps in regions N1, N2, and N3 are all located near the 
tip of the lobe.
Such an arc-like spatial distribution of molecular gas seems somewhat
similar to the simulational 
result with a large filling factor of HI clumps 
\citep{2017ApJ...836..213A}.

\subsection{Molecular line profiles}
\label{kinematic}
We have inspected the CO line profiles toward N2 and N3.
For clumps with $\VLSR<+40\km\ps$,
neither an asymmetric broad-line profile of the
\twCO\ (\otz) nor a high \twCO\ \tto/\otz\ line ratio
has been found.
For clumps around $+53\km\ps$, we made grids of \twCO\ (\otz),
\twCO\ (\tto), and \thCO\ (\otz) spectra in the velocity range
$+47$ to $+61\km\ps$ and $+49$ to $+61\km\ps$ 
(part of N2 and N3 are shown in 
Figure~\ref{linegridn1} and Figure~\ref{linegridn2}, respectively).
There are some common characteristics 
of the spectra of the clumps in regions N2 and N3.

(1) 
The kinetic temperatures $T_{\rm c,k}$ 
estimated from the peak of the main-beam temperatures
can reach 28/30\,K for N2/N3 (see Table~\ref{parameter}),
which are significantly higher than all
those ($\sim 10$\,K, similar to the typical temperature for interstellar 
MCs) at other LSR velocities
(see Figure~\ref{tkin} for a comparison with the clumps at $\sim +26$
and $\sim +8\km\ps$).
Here we consider that $T_{\rm c,k}$ is equal to the excitation temperature
$T_{\rm ex}$ under the assumption of local thermodynamic equilibrium
(LTE). $T_{\rm ex}$ is calculated from the main-beam temperature
$T_{\rm mb}$ using the equations
$T_{\rm ex}=5.53/\{\ln[1+5.53/(T_{\rm mb,10}+0.84)]\}$\,K for \twCO\ (\otz)
and $T_{\rm ex}=11.06/\{\ln[1+11.06/(T_{\rm mb,21}+0.2)]\}$\,K for 
\twCO\ (\tto).

(2) The \twCO\ \tto/\otz\ line ratios,
$R_{21/10} =T_{\rm MB,21}/T_{\rm MB,10}$,
of the clumps are high.
The $R_{21/10}$ maps and the $1\sigma$ error maps for the LSR velocity 
intervals $+50$ to $+60\km\ps$, $+24$ to $+28\km\ps$, and $+6$ to $+10\km\ps$
are shown in Figure~\ref{ratio}.
The line ratios are $> 0.9$ for most of the clumps at $\sim +53\km\ps$.
As a comparison, the ratios for the clumps at $\sim +26$ and
$\sim +8\km\ps$ 
are $\la 0.8$ (see the middle left and bottom left panels 
in Figure~\ref{ratio}).

\begin{figure}[t]
\centering
\includegraphics[scale=.67]{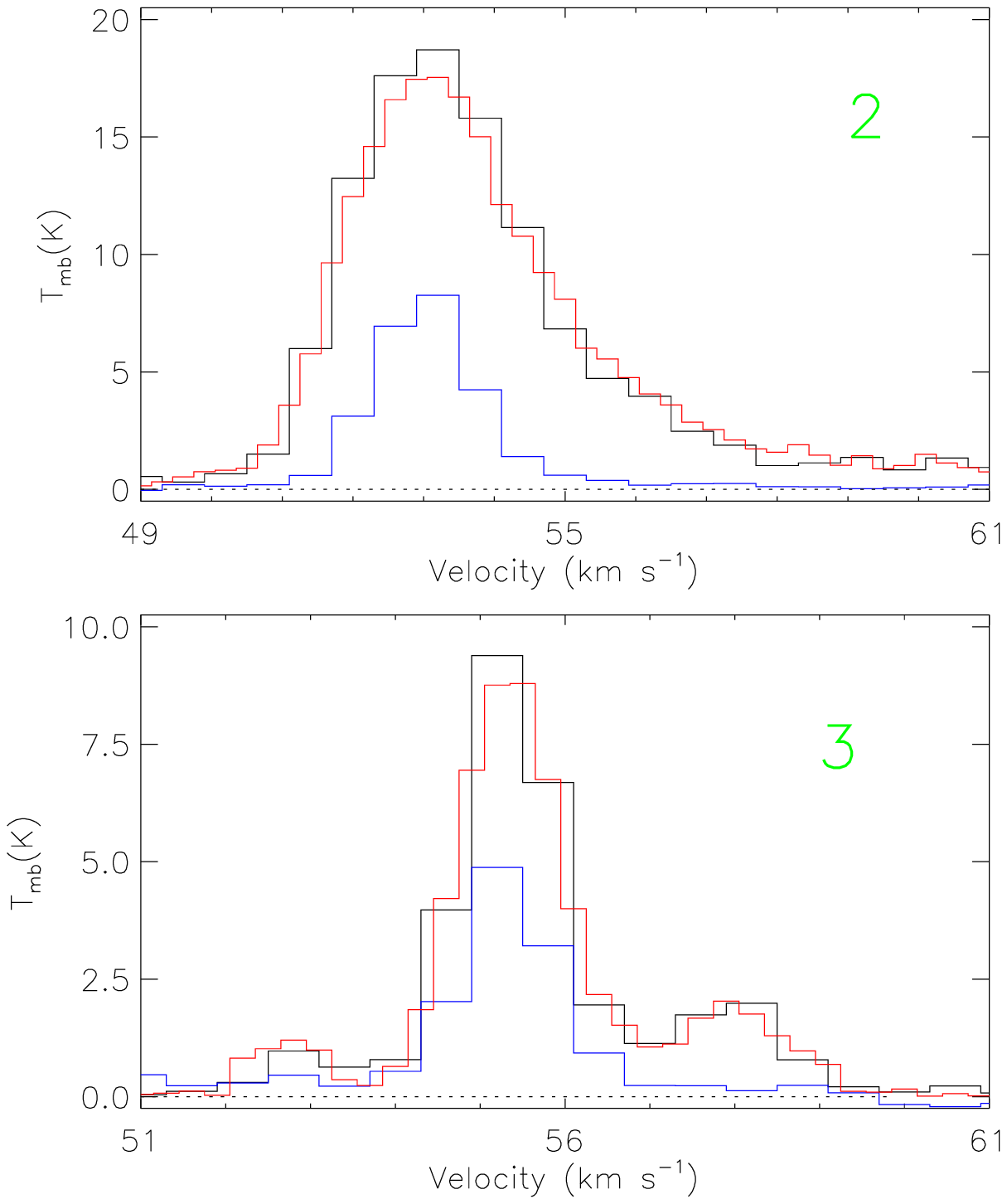}
\caption{\footnotesize Averaged spectra of regions ``2" (top panel) and ``3"
(bottom panel) in the velocity ranges
$+49$ to $+61\km\ps$ and $+51$ to $+61\km\ps$, 
respectively, obtained from the IRAM observation.
The two regions are defined in Figure~\ref{linegridn1}
and \ref{linegridn2}.
The black thick lines denote the \twCO\ (\otz) spectra, 
the red lines represent \twCO\ (\tto), the blue lines
represent \thCO\ (\otz), and the dotted lines
represent the 0\,K main-beam temperature. 
The \thCO\ (\otz) spectrum in the bottom panel has been multiplied
by five for better visibility.
\label{asymmetry}}
\end{figure}

(3) Both N2 and N3 show some asymmetric broad profiles of the \twCO\
line.
Figure~\ref{asymmetry} shows the average line profiles
delineated by the green rectangles labeled ``2" 
and ``3" in Figure~\ref{linegridn1} and \ref{linegridn2}.
The red (right) wings of the \twCO\ lines appear to be broadened at 
$\sim +54$ to $+59\km\ps$ 
and $\sim +56$ to $+59\km\ps$ for regions ``2" and ``3", respectively.
The one-side widths $\sim 5$ and $\sim 3\km\ps$ of the broad right (red) wings 
are notably larger than the half-widths of the FWHMs
($\sim 1.5$ and $\sim 0.9\km\ps$, Table~\ref{parameter})
of the main bodies of the velocity components, respectively.
There is no sign that these line wings
could be contaminated from their surrounding pixels,
somewhat similar to the \twCO\ spectra detected in the southwestern
edge of SNR~CTB\,87 \citep{2018ApJ...859..173L}.
That is, there are strong peaks at $\sim +53$/$\sim +55\km\ps$
for both \twCO\ and \thCO\ in regions ``2"/``3", while the broad
red wings of \twCO\ extend to $+59\km\ps$ without 
significant \thCO\ counterparts, and their surroundings
are weak in the velocity range where the \twCO\ red wings are.
Such asymmetric/one-sided broadened \twCO\ line profiles are present 
in many SNR-MC interaction systems, like 
Kes\,75 \citep{2009ApJ...694..376S}, 3C\,397 \citep{2010ApJ...712.1147J}, 
Kes\,78 \citep{2011ApJ...743....4Z}, 3C\,396 \citep{2011ApJ...727...43S},
and CTB\,87 \citep{2018ApJ...859..173L}.
Usually, the asymmetric part deviating from the main Gaussian in a 
\twCO\ profile is believed to be a broadened part if there is 
little \thCO\ line feature at the corresponding LSR velocity.
Because \thCO\ emission, whose optical depth is smaller than 1
in most cases (i.e., optically thin, which is satisfied in this
work, as is given in Table~\ref{parameter}),
traces the quiescent gas,
the lack of a significant \thCO\ feature 
at the velocity indicates that the deviating part very likely to 
represents the disturbed gas due to external interaction 
\citep[e.g.,][]{2011ApJ...727...43S, 2011ApJ...743....4Z, 
2016ApJ...831..192Z}.
Therefore,
the broad \twCO\ red wings from regions ``2" and ``3" 
are highly likely to result from Doppler broadening
of the $\sim +53$/$\sim +55\km\ps$ lines
and thus could represent a kinematic signature 
of the perturbation of 
the $\sim +53$/$\sim +55\km\ps$ MC clumps
by external impacts, especially from jet-related gas.

\begin{figure*}
\centering
\includegraphics[scale=.45]{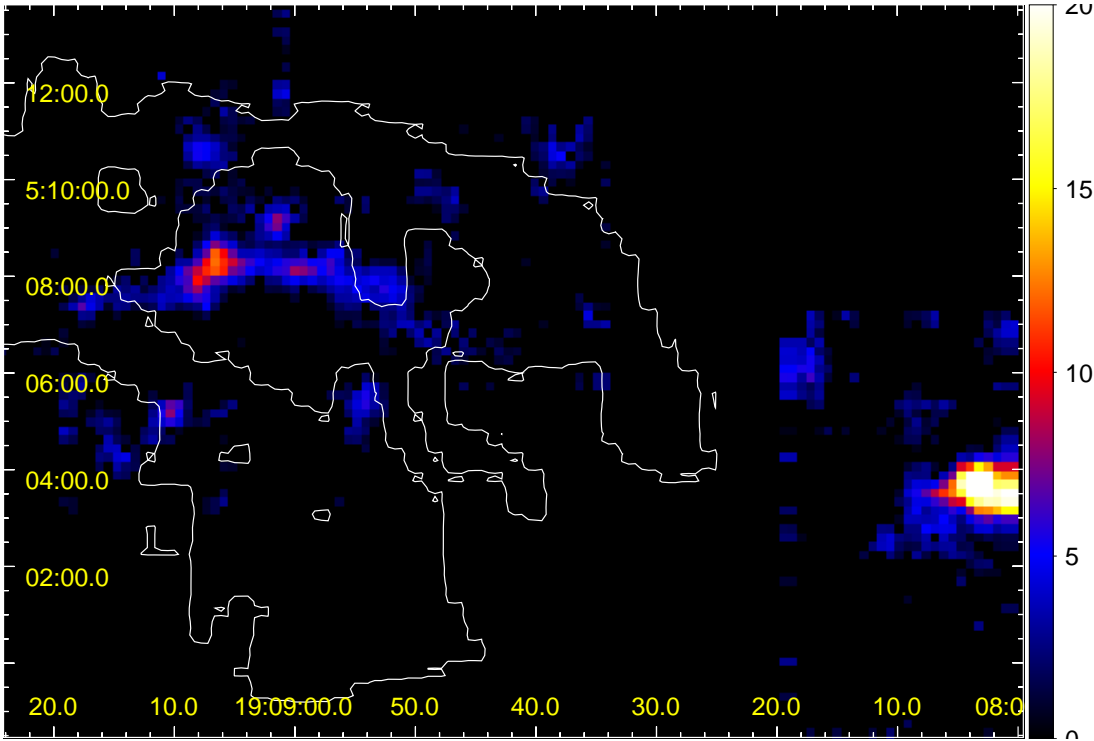}
\caption{\footnotesize $\NHH$ distribution of regions N2 and N3 in the velocity 
range of $+50$ to $+60\km\ps$ in units of $10^{21}\psc$
obtained from the IRAM observation.
The radio contour is the same as that in Figure~\ref{channelmap1}.
\label{cddistribution}}
\end{figure*}

Table~\ref{parameter} summarizes the observed and derived parameters
for the two clumps in regions N2 and N3 as observed in the three CO lines
at $+50$ to $+60\km\ps$.
We then estimated the distribution of the column density
$\NHH$ using \thCO\ lines, 
under the assumption of LTE for the MCs, optically thick
condition for the \twCO\ (\otz) line, and optically thin 
condition for the \thCO\ (\otz) line.
The excitation temperatures of the two clumps are assumed to be
$T_{\rm ex}=28$\,K for N2 and $T_{\rm ex}=31$\,K for N3 
(see Table~\ref{parameter}).
Here we have used the conversion relation for the 
molecular column density of 
$N$(H$_2$)\,$\approx 7\times 10^5 N$~(\thCO) \citep{1982ApJ...262..590F}.
The estimated $\NHH$ distribution at $+50$ to $+60\km\ps$
is shown in Figure~\ref{cddistribution}, with maximum values of
$\sim 2.3\times 10^{22}\psc$ in N2 and 
$\sim 1.2\times 10^{22}\psc$ in N3
(however, note that the column density could be 
overestimated by a factor of up to 4 for the \thCO;
see \citealt{2013tra..book.....W}).
The H$_2$ masses and mean H$_2$ densities in the field of view 
are also estimated and summarized in Table~\ref{parameter},
when parameterizing the distance to the MC as
$d = 3.5\du\kpc$
\citep{2008PASJ...60..715Y}.

\begin{table*}[t]
\centering
\footnotesize
\begin{threeparttable}
\caption{\footnotesize Information of CN (\cnj) \label{cnposition}}
\begin{tabular}{ccccccccc}
\hline \\
Position & R.A.\tnote{a} &
Decl.\tnote{a}
& Size \tnote{b} & $T_{\rm max}$\tnote{c} &
Line Width\tnote{d} &  
N(CN)\tnote{e} & N$\HH$ & N(CN)/N$\HH$\\
 & & & & (K) & ($\km\ps$) & ($\times 10^{14}\psc$) & 
($\times 10^{21}\psc$) & ($\times 10^{-8}$)\\
\hline\\
1 & 19:09:14.95 & 05:04:26.0 & $33''\times 33''$ & 0.78 & 
1.0 & 1.7 & 2.8 & 6.1 \\
2 & 19:09:09.80 & 05:05:21.0 & $11''\times 33''$ & 0.48 &
1.2 & 1.2 & 5.5 & 2.2 \\
3 & 19:08:54.34 & 05:05:21.0 & $33''\times 22''$ & 0.45 &
1.4 & 1.3 & 3.8 & 3.4 \\
4 & 19:09:01.80 & 05:09:12.0 & $44''\times 55''$ & 0.46 &
1.1 & 1.1 & 3.9 & 2.8 \\
5 & 19:09:08.33 & 05:07:55.0 & $22''\times 33''$ & 0.45 &
1.1 & 1.0 & 8.3 & 1.2 \\
6 & 19:08:02.97 & 05:03:42.3 & $11''\times 33''$ & 0.77 &
1.6 & 3.2 & 19.5 & 1.6 \\
\hline
\end{tabular}
\begin{tablenotes}
\item[a] \footnotesize The central positions of the six regions
\item[b] The CN (\cnj) emission lines are averaged within
these region sizes.
\item[c] The maximum CN (\cnj) emission brightness 
temperature in each region
\item[d] The FWHM of each CN (\cnj) emission line
\item[e] The column density of CN is estimated with an assumed 
molecular density of $10^5\pcc$ and kinetic temperature of 28\,K
(N2, position 6) or 31\,K (N3, position 1--5).
\end{tablenotes}
\end{threeparttable}
\end{table*}

\begin{figure}
\centering
\includegraphics[scale=.55]{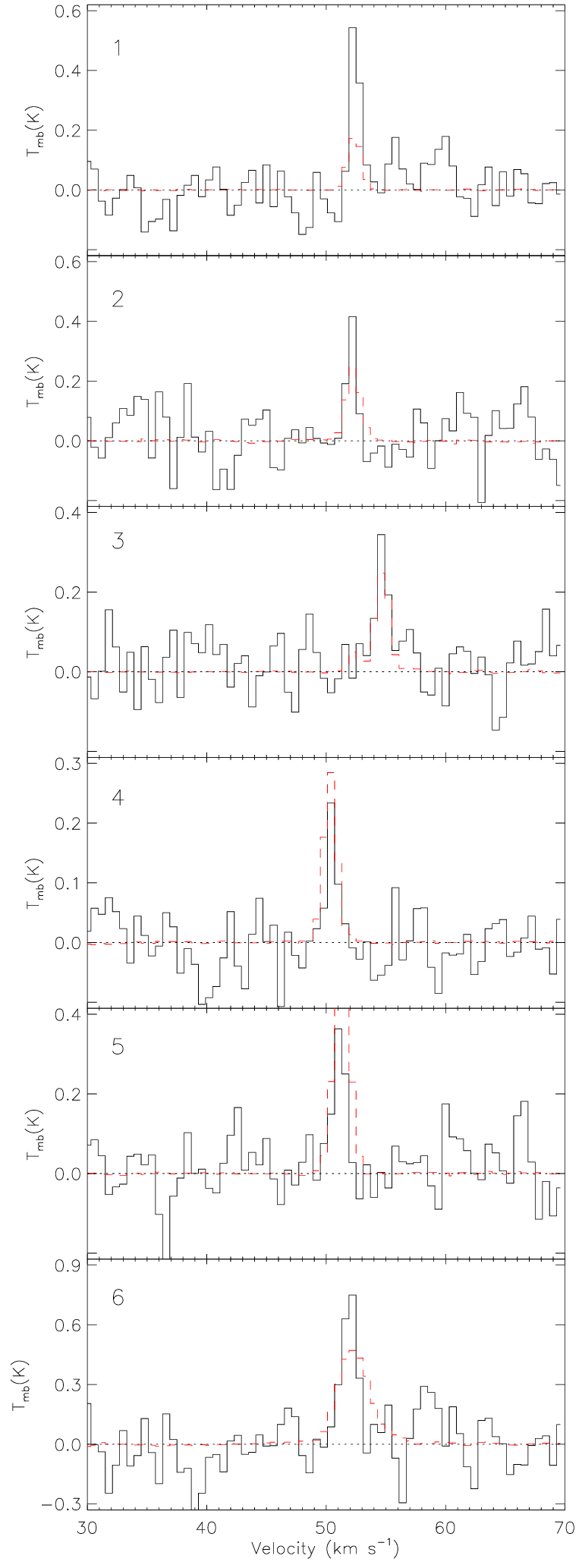}
\caption{\footnotesize IRAM CN (\cnj) spectra extracted from the regions labeled
in the upper left panel of Figure~\ref{multiwave},
in velocity range $+30$ to $+70\km\ps$.
The dotted lines represent the 0\,K main-beam temperature.
The position information of these regions are summarized
in Table~\ref{cnposition}.
As a comparison, \twCO\ (\otz) spectra extracted from 
same regions are also plotted (the red dashed lines,
which have been multiplied by a factor of 0.02).
\label{cnlines}
}
\end{figure}

\begin{figure}[t]
\centering
\includegraphics[scale=.5]{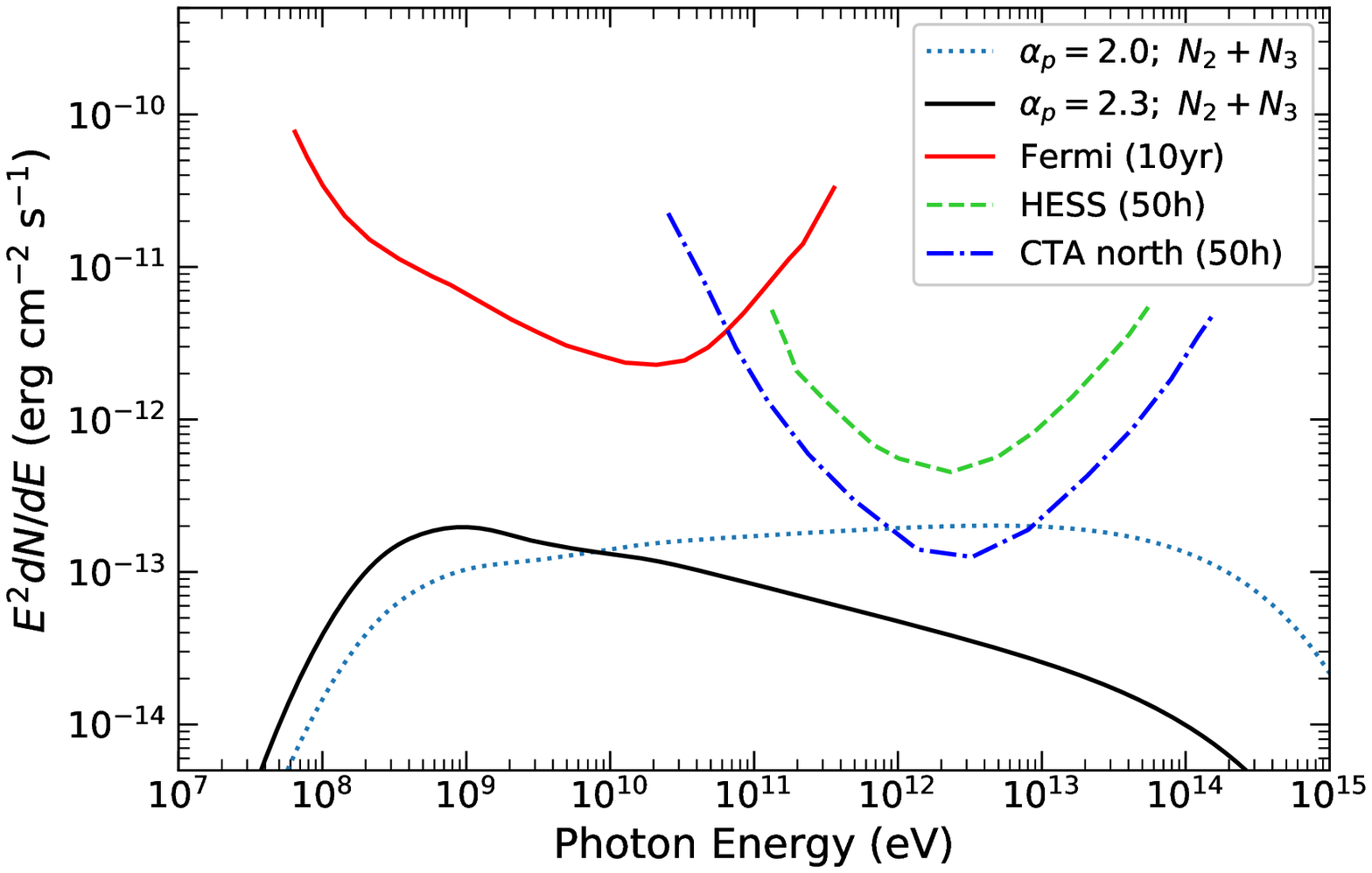}
\caption{\footnotesize
Expected gamma-ray fluxes from clumps in regions N2 and N3,
with assumed proton indexes of 2.0
(dotted line) and 2.3 (black line). The sensitivity of 
Fermi LAT in 10\,yr (red line, \citealt{2013APh....43..348F}), 
HESS in 50\,hr (dashed line, \citealt{2016NPPP..279..166D}), 
and CTA north in 50\,hr (dashed-dotted line, \citealt{2016NPPP..279..166D})
are overlaid.
\label{gamma}
}
\end{figure}

\subsection{Emission lines of CN at $+50$ to $+58\km\ps$} \label{cnsection}
We detected emission lines of CN ($J$=3/2--1/2) at six small regions
(labeled with white crosses in the upper left panel of 
Figure~\ref{multiwave}, each of only several pixels).
The CN emission at $+50$ to $+58\km\ps$ could arise from
denser part of the MCs with density of 
order $10^5\pcc$ \citep{1975ApJ...198...71T},
which is much higher than the average
densities of the two clumps estimated from the CO lines
(of order $10^3\pcc$, see Table~\ref{parameter}).
The position information and the size of the six regions
are summarized in Table~\ref{cnposition}.
Figure~\ref{cnlines} shows the CN (\cnj) spectra from the 
six regions. 

Though there are five CN (N=1-0, J=3/2-1/2) hyperfine splitting
lines in the frequency 113.488--113.521\,GHz range,
we suggest the CN line we observed in this work to be
the CN (\cnj) hyperfine line, based on the observational result 
that the CN line we observed was coincident with the CO 
spectra in the velocity interval +50 to $+60\km\ps$ 
(see Figure~\ref{cnlines}) if we set the reference frequency
to be CN (\cnj) of 113.491\,GHz.
Also, CN (\cnj) has the strongest relative intensity
among the hyperfine lines \citep{1975ApJ...198...71T}.
The second strongest hyperfine line, CN ($J$=3/2--1/2, 
$F$=3/2--1/2) at 113.488\,GHz, which is $\sim 7\km\ps$
away from the CN (\cnj), has not been detected 
within the noise level.

The column densities of CN are estimated with radiative 
transfer calculations of the CN (\cnj) lines using the RADEX code
under the large velocity gradient approximation \citep{2007A&A...468..627V}.
The input parameters, including the kinetic temperature, the
line width, and the brightness temperature, are taken as
the estimated excitation temperature of N2 and N3 
(see Table~\ref{parameter}),
the maximum CN (\cnj) emission point, the fitted FWHM of the CN (\cnj)
emission line, respectively.
The best-fitting results of the CN column densities are shown
in Table~\ref{cnposition}.

In Table~\ref{cnposition}, we also list the column densities of 
$\NHH$ and the CN abundances, $N$(CN)/$\NHH$,
toward each position where the CN emission is detected.
The CN abundances are in the range $\sim (1$--$6)\times10^{-8}$,
which is about an order of magnitude higher than that in 
the interstellar MC 
\citep[e.g.,][]{1987ApJ...315..621B, 2014ApJ...788....4W}
and that in the MC an SNR interacts with
\citep[e.g.,][]{1992ApJ...399..114T}.
The derived CN abundances 
are consistent with the abundance ($\sim1.4$--$4.1\E{-8}$)
estimated for an MC that interacts with cosmic rays (CRs) on timescales of 
$10^4$--$10^5$\,yr \citep{2018ApJ...868...40A}.

\section{Discussion} \label{discussion}

Our small-scale CO-line observation shows that the two clumps
at LSR velocity around $+53\km\ps$
near the tip of the western lobe of the SS\,433/\snr\ system
are spatially correspondent with multiwavelength
local features of the system.
Toward the two clumps, asymmetric broad-line profiles of the
\twCO\ lines are obtained, providing kinematic evidence 
of the association between the clumps and the jet-related gas.
The \twCO\ \tto/\otz\ line ratios ($\ga 0.9$), as well as the kinetic 
temperatures ($\sim 30\,{\rm K}$) of the
two clumps, are distinctively higher than all those of the MCs at
other LSR velocities along the same LOS, which may be physical
signatures of the association.
Some physical issues in this scenario are discussed below.

\subsection{The anticorrelation between the radio emission and the clump
in region N3}
\label{anticor}
With the properties favoring an interaction between the 
clumps at $\sim +53\km\ps$ and the western lobe of SS\,433
(i.e., the asymmetrical \twCO\ lines,
high \twCO\,\tto/\otz\ line ratios, and the high molecular gas 
temperatures) obtained, the
anticorrelation between the clump in region N3 and the radio continuum
(as shown in Figure~\ref{multiwave}) are not likely 
due to a chance projection effect.
It could hardly result from extinction by the clump 
located in the foreground of SS\,433,
because either the electron scattering or the free-free absorption
is negligible for the extinction of the radio emission.
The optical depth of electron scattering can be estimated as
$\tau_{\rm es} = \sigma_e n_e l$,
where $\sigma_e = 6.65\times10^{-25}\cm ^{-2}$ is the total
cross section of an electron, $n_e$ is the electron density
($\sim0.05\pcc$) for a clump with a density of $\sim 10^3\pcc$
(Table~\ref{parameter})
and a CR ionization rate of 
$\zeta\HH \sim 5\times 10^{-16}\ps$
(see \S \ref{thermal})
and $l$ ($\sim 1\,{\rm pc}$) is the length of the clump along the 
line of sight (LOS). 
We thus have $\tau_{\rm es}\sim 1\times 10^{-7} \ll 1$.
The optical depth of free-free absorption by electrons
can be expressed as
$\tau_{\rm ff} = k_\nu l$, where
$k_\nu = 0.1731[1+0.13\log(T_{\rm c,k}^{3/2}/\nu)]
({\rm Z}^2 n_e n_i)/(T_{\rm c,k}^{3/2}\nu ^2){\rm cm}^{-1}$
\citep{1978ppim.book.....S},
in which $Z$ is the charge of ions ($\sim1$),
and $n_i$ is the ion density (very similar to $n_e$).
The optical depth of free-free absorption can thus be estimated
to be $\tau_{\rm ff} \sim 3\times 10^{-7} \ll 1$ for
$T_{\rm c,k}\sim 30$\,K 
and $\nu=1.4$\,GHz.
Therefore, 
the anticorrelation between the CO emission and the radio continuum
is not caused by the molecular gas extinction, similar to the case
in SNR CTB\,109 \citep[e.g.,][]{1987A&A...184..279T,
1992ApJ...388..127W}.
It hence suggests a scenario in which the clump in region N3 
is embedded in the radio-emitting gas in the western lobe of SS\,433.

\subsection{Evaporation of the clump}
We have indicated that the clump in region N3 is embedded in the western lobe. 
This scenario is consistent with the location of this clump, 
which seems to be in a void of X-ray emission (\S~\ref{spatial}).
If it implies that the clump is surrounded by the X-ray-emitting hot gas
of the western lobe, the clump seems to survive the heating of the gas.

For the case in which the X-rays of the lobe 
are dominated by thermal bremsstrahlung,
the gas temperature is $\sim 3.3\,{\rm keV}$
or $\sim 4\times 10^7\,{\rm K}$ \citep{1997ApJ...483..868S}.
For simplicity, we only estimate the evaporative effect on the
clump due to saturated thermal conduction from the hot gas.
As a result of heating on the clump surface
by inward heat flux from the high-temperature,
low-density environment,
the cold clump could gradually evaporate
\citep{1977ApJ...211..135C}.
It is essential to compare the evaporation
timescale of the clump in N3 with the age of the SS\,433 system.
The saturated mass-loss rate is given by
$\dot{m} = 3.25\times 10^{18}n_{\rm h}T_{\rm h}^{1/2}R_{\rm pc}^2\phi
F(\sigma_0){\rm g}\ps$
\citep{1977ApJ...211..135C},
where $n_{\rm h}$ and $T_{\rm h}$ are the atomic density and temperature
for the hot medium, 
where $R_{\rm pc}$ is the radius of the 
clump in units of parsecs, $\phi$ is a factor of 
order unity to equate the saturated heat flux and the 
pressure of the medium times the isothermal sound speed,
and $F(\sigma_0)$ is a function with a value $\sim 8$ for
$\sigma_0$, the saturation parameter,
around $\sim 20$ under the physical condition of the hot gas considered here.
The evaporation timescale of the clump in region N3
can be estimated as
\begin{equation}
\begin{split}
\tau_{\rm ev} &\sim M/\dot{m} \sim 8\times 10^5\du^2 
R_{\rm pc}^{-2} \phi ^{-1} \\
&\left( \frac{n_{\rm h}}{0.3 \pcc}\right )^{-1} 
\left(\frac{T_{\rm h}}{4\times 10^7\,{\rm K}}\right)^{-1/2} 
\left(\frac{F(\sigma_0)}{8}\right)^{-1}\,{\rm yr},
\end{split}
\end{equation}
where $M$ is the mass of the clump ($\sim500\du^2 M_{\odot}$ for that 
in N3; Table~\ref{parameter}),
and a number density $n_{\rm h}\sim 0.3\pcc$ given in
the thermal bremsstrahlung model
\citep{1997ApJ...483..868S} is adopted as a reference value.
By contrast, it is suggested that the age of SNR\,W50 is 
$\sim 10^5\,{\rm yr}$ and the age of the SS\,433 jets is
$< 2.7\times 10^4\,{\rm yr}$
\citep{2017A&A...599A..77P}, both of which are much smaller than 
the evaporation timescale. This means that
evaporation of the cold gas is far from exhausting the embedded clump.

\subsection{The heating source of the clumps}\label{thermal}

The \twCO\ \tto/\otz\ line ratios are as high as
0.9--1 for the clumps in regions N2 and N3
(\S\,\ref{kinematic}), implying that the gas 
temperatures of the clumps could be evidently higher than
the typical temperature of interstellar molecular gas
($\sim 10$\,K).
The line ratio is given by
$R_{21/10}
=f_{21}\nu_{21}[J_{\nu_{21}}(T_{\rm ex})-J_{\nu_{21}}(T_{\rm bg})]
[1-e^{-\tau_{21}}]
/\{
f_{10}\nu_{10}[J_{\nu_{10}}(T_{\rm ex})-J_{\nu_{10}}(T_{\rm bg})]
[1-e^{-\tau_{10}}]\}$,
where $T_{{\rm MB},ji}$, $f_{ji}$, $\nu_{ji}$, and $\tau_{ji}$ are the 
main-beam temperature, filling factor, frequency, and optical depth
of the \twCO\ $J=j$--$i$ line emission, respectively,
$J_{\nu_{ji}}(T)\equiv(e^{h\nu_{ji}/kT}-1)^{-1}$, and $T_{\rm bg}\sim 
2.73$\,K is the background temperature.
Since the clumps are optically thick ($\tau_{ji}>1$) and diffuse 
($f_{ji}=1$) for \twCO\ \tto\ and \otz\ line emission,
the line ratio can be approximated as
\begin{equation}
R_{21/10}\approx 2\times[J_{\nu_{21}}(T_{\rm ex})-J_{\nu_{21}}
(T_{\rm bg})]/[J_{\nu_{10}}(T_{\rm ex})-J_{\nu_{10}}(T_{\rm bg})],
\end{equation}
which is 
$\ga0.9$ for $T_{\rm ex}\ga 20$\,K. 
Therefore, the line ratios 
observed in the 
two clumps appear to be consistent with the obtained kinetic temperatures
(28/30\,K, \S\,\ref{kinematic}).
Also, such line ratio and temperature are clearly
higher than all those ($\sim$ 0.8 and $\sim 10$\,K) 
of the molecular gas
at other LSR velocities along the same LOS 
(see the comparisons in Figures~\ref{tkin} and \ref{ratio}).
There should be heating source(s) for the clumps, with possible candidates
including environmental X-rays, external disturbance, and CRs.
The heat flux from the surrounding hot gas is consumed by the
gas evaporation on the clump surface.

The environmental diffusive X-ray emission plays
an insignificant role in heating the clumps,
although the X-rays could heat the molecular gas in the 
far-UV-shielded regions (i.e., regions beneath the surface
of the MCs with column density larger than $10^{21}\psc$;
see, e.g., \citealt{2018ApJ...865....6Z}).
\cite{1997ApJ...483..868S} gives the X-ray luminosity of a
circular region ``w2", the brightest portion in the lobe,
near the clump in region N3:
$L_{\rm X}\la
2\times 10^{34}(d/5.5\,{\rm kpc})^2\erg\ps$ in 0.1--2.4\,keV.
The angular radius of the circle is ${\cal R}=3.75'$, which can be translated 
to ${\cal R} \sim 3.9\du\parsec$.
The X-ray flux around the sphere indicated by this circle region
$F_{\rm X}= L_{\rm X}/(4\pi {\cal R}^2)
\la 5\times 10^{-6}\erg\psc\ps$
is used as an upper limit of the X-ray flux
at the clump in N3.
The X-ray heating rate can be approximated as
\citep{1996ApJ...466..561M}
$\Gamma_{\rm X} = 1.4\times 10^{-26}
[n\HH/10^3\pcc][F_{\rm X}/(5\times 10^{-6}\erg\psc\ps)]
(N_{\rm H}/10^{22}\psc)^{-0.9}\erg\pcc\ps$,
where $N_{\rm H}$ is the column density of hydrogen 
attenuating the X-ray flux.
On the other hand, the cooling of MCs with kinetic temperature
of dozens of Kelvin and density of $\sim 10^3 \pcc$ 
is dominated by molecular lines, 
and the cooling rate can be approximated by
\citep{2001ApJ...557..736G} 
$\Lambda_{\rm gas} \sim 1.5\times 10^{-23}
(T_{\rm c,k}/30 {\rm K})^{2.4}$
$[(dv/dr)/(1\km \ps {\rm pc}^{-1})]\erg \pcc \ps$,
where $dv/dr$ is the velocity gradient of the MCs.
The X-ray heating rate is about three orders of magnitude
smaller than the cooling rate, indicating that 
the X-ray heating is actually negligible. 

The obtained broadened \twCO\ lines can be ascribed to an external
disturbance, which could have a heating effect on the clumps.
However, it cannot play an important role in heating the clumps.
The transmission velocity of the disturbance should be comparable to the 
redshifted \twCO\ line broadening ($\sim 3$--$5\km\ps$).
The timescale for the disturbance to cross the clump with a typical size of
$\sim1$\,pc is about $(2$--$3)\times 10^5\,{\rm yr}$,
which is an order of magnitude larger than the age of the jets ($<2.7\times 10^4
\,{\rm yr}$, \citealt{2017A&A...599A..77P}).

CRs could penetrate deeply into the far-ultraviolate shielded
regions of MCs and ionize the molecules. 
The ionization process can provide a heating rate
\citep{1978ApJ...222..881G}
$\Gamma_{\rm CR}=\zeta \HH \Delta Q n\HH
\sim 6.4\times 10^{-25} [n\HH /10^3 \pcc]
[\zeta \HH /(2\times 10^{-17}\ps)] \erg \pcc \ps$,
where $\zeta\HH$ is the ionization rate of H$_2$ by CR
and $\Delta Q$ is the energy deposited as heat 
per ionization.
Here a mean value of 20\,eV is adopted for $\Delta Q$,
though it varies with CR energy and gas composition
\citep[e.g.,][]{1978ApJ...222..881G,2012ApJ...756..157G}.
To maintain a temperature $T_{\rm c,k}\sim 30$\,K via 
thermal equilibrium $\Gamma_{\rm CR} = \Lambda_{\rm gas}$,
the ionization rate should be
\begin{equation}
\begin{split}
\zeta\HH & \sim 4.7\times 10^{-16} 
\left(\frac{T_{\rm c,k}}{30 {\rm K}}\right)^{2.4} \\
&\left(\frac{n\HH}{10^3 \pcc}\right )^{-1}
\left(\frac{dv/dr}{1\km\ps{\rm pc}^{-1}}\right)\ps.
\end{split}
\end{equation}
This rate appears somewhat larger than the background CR ionization rate
ranging from $\sim 2\times 10^{-17}\ps$
\citep{1978ApJ...222..881G}
to $\sim 2\times 10^{-16}\ps$
\citep[e.g.,][]{2012ApJ...745...91I}.
It is, however, not surprising because the energy density
of CRs may be expected to be higher at locations (like N3 and N2)
inside or close to the
boundary of an SNR (i.e., \snr)
and near the energetic jets than the Galactic mean.

In the scenario the two observed clumps are associated
with the \snr/SS\,433 system,
the energy expended to maintain the high molecular temperature
on a timescale $\tau$ is
$\Gamma_{\rm CR} [M/\rho\HH] \tau$,
where $\rho\HH$ is the mass density of the molecular gas;
for the clump in N3 and adopting $\tau$ as the age of SNR\,\snr,
this energy is
$\sim1.4\times10^{46} (M/500M_\odot)
(T_{\rm c,k}/30 {\rm K})^{2.4}$
$[n\HH /(10^3 \pcc)]^{-1}$
$[(dv/dr)/(1\km\ps{\rm pc}^{-1})]$
$(\tau/10^5\,{\rm yr})\,{\rm erg}$.
A canonical energy for the CRs accelerated by the SNR shock,
about 10\% of the supernova explosion energy $10^{51}$\,erg,
together with the energy that may be potentially
converted from the SS\,433 jets, is a few
orders of magnitude larger than the
energy needed to maintain the thermal equilibrium in the clumps.

\subsection{Gamma-Ray emission from the clumps?}
The CRs may produce considerable gamma-rays by pp interactions,
a process which has also been discussed toward the central part 
of SS\,433
\citep[e.g.,][]{2008MNRAS.387.1745R}.
To calculate the 
gamma-ray flux, one needs to know the energy density ($w$) of CRs in the
clumps, which 
can be estimated by scaling the ionization rate obtained in this paper to 
that caused by the background CRs. 
The ionization rate for the background CRs above 10\,MeV is estimated to be
$\sim 2\times10^{-17}\ps$ \citep{1973PASJ...25..327T,
1978ApJ...222..881G}. 
From Eq.\ A1 in \citet{1973PASJ...25..327T}, we obtain the energy density 
of the background CRs above 1\,GeV 
$w_0(>1\ {\rm GeV})\sim0.4 \,{\rm eV}\pcc$. 
Thus, according to the ionization rate obtained, we have
$w(>1\ {\rm GeV})\sim20w_0\sim8\,{\rm eV}\pcc$ as the energy density of 
the CRs that heat and maintain the high temperature of the molecular clumps.
We further assume the CR distribution has a power-law form with a proton 
index $\alpha_p$ and a maximum energy of 3 PeV.
Then the $\pi^0$-decay gamma-ray emissivity per H-atom, 
$q_{\gamma}(E_{\gamma})$, can be obtained according to the formulae given in 
\citet{2014PhRvD..90l3014K}.
The expected gamma-ray flux from the two clumps is
$(M_{cl}/m_p)\times (q_{\gamma}(E_{\gamma})/4\pi d^2)$, which is shown in 
Figure~\ref{gamma}.
Here, we consider two cases with different proton index: $\alpha_p=2.0$ and 
$\alpha_p=2.3$. For both cases the gamma-ray flux estimated is lower than the 
sensitivity of current telescopes, Fermi and HESS in the GeV and TeV band, 
respectively, which is consistent with the report that no evidence of 
gamma-ray emission from the jet termination regions has been found
between a few hundred GeV and a few TeV
\citep{2018A&A...612A..14M}.
For $\alpha_p=2.0$, the gamma-ray flux seems to be detectable with CTA, 
the next-generation TeV telescope.
However, if the transport of CRs to the clumps is governed by 
diffusion, 
the proton index should be $\alpha_p=\alpha_p'+\delta$, where $\alpha_p'$ 
is the proton index at the acceleration site and $\delta$ is the index of 
the diffusion coefficient $D(E)\propto E^{\delta}$.
If one takes $\alpha_p'=2.0$ as predicted by the standard diffusive shock 
acceleration theory and the typical value of $\delta\sim0.3$--0.7 for the 
diffusion coefficient in ISM, the proton index in the clumps 
should be greater 
than 2.3. Therefore, the expected gamma-ray emissions are even undetectable
with a next-generation telescope.

\subsection{Clump formation?}
The clumps in regions N2 and N3, as well as in N1, are all located 
near the tip of the western X-ray or radio lobe (\S~\ref{spatial}).
An MHD simulation suggests that MCs can form by the jet-HI clump interaction:
when the filling factor of the HI clumps the jet interacts with 
is large, arc-like distributed MCs with thicknesses of $\sim 1$\,pc form
\citep{2017ApJ...836..213A}. Compared with the observed arc-like spatial 
distribution of the clumps in this work,
the simulation could imply that the filling factor of the HI
clumps is large to the west of SS\,433,
under the scenario that the clumps are potentially produced
by jet-HI interaction.
This is in general consistent with the environment of the clumps
in the west of SS\,433, where the Galactic latitude is low,
and the HI gas is dense.
In addition, the high CN abundances found in some positions in
the molecular clumps are consistent
with model predictions of MCs which interact with CRs on timescales of 
$10^4$--$10^5$\,yr (\S~\ref{cnsection}).
This timescale is comparable to the age of the SS\,433 jet 
$<2.7\times 10^4$\,yr \citep{2017A&A...599A..77P}.
However, to confirm if the clumps are induced by jet-HI interaction,
further investigations are needed.

\section{Summary} \label{summarize}
We perform a small-scale CO-line observation using the IRAM 30m 
telescope toward two regions (N2 and N3), near the western tip of the X-ray
and radio emission, that are potentially the result of an 
interaction of the jet of SS\,433 and the surrounding medium.
The observational and physical properties of the molecular clumps
in the two regions are summarized below.
\begin{enumerate}

\item{The CO observation shows that two clumps in the two regions
at around $\VLSR\sim +53\km\ps$
are spatially correlated with dense gas indicated by
two patches of enhanced mid-IR emission.
The clump in region N3 appears to be embedded in 
a void of diffuse radio emission,
while the western edge of the X-ray emission is 
finely enclosed by the mid-IR emission.
The negligible optical depth of the clump
indicates that the radio void is not likely due to an
extinction by the clump located in the foreground 
}

\item{
Asymmetric broad-line profiles of the $\sim +53\km\ps$
\twCO\ lines are obtained, providing kinematic evidence 
of the association between the clumps and jet-related gas.
The \twCO\ \tto/\otz\ line ratios ($\ga$ 0.9) and the kinetic temperatures
($\sim 30$\,K) of the two clumps are distinctively higher than all those
of the MCs at other LSR velocities along the same LOS, which may be
physical signatures of the association.
}

\item{We also detect emission lines of CN ($J=3/1$--$1/2$) at
six positions in the two clumps, and the derived CN abundances
are about an order
of magnitude higher than those in the interstellar MCs
and in the MCs with which SNRs interact.}

\item{
We show that the clump in region N3 can survive thermal heating 
if it is surrounded by hot gas, with an evaporation timescale much 
larger than the age of SNR\,\snr. We also show that the thermal 
equilibrium in the high-temperature clumps can be maintained by 
heating of the penetrating environmental CRs.
}
\end{enumerate}

\begin{acknowledgements}
We are grateful to Yang Su for a discussion of the \snr/SS\,433
environment and MC physics. We thank the staff of the IRAM 30\,m 
observatory for help during the observation. 
We acknowledge the MWISP project.
This work is supported by the National Key R\&D Program of China under grants 
2017YFA0402600, 2015CB857100 and the NSFC under grants 11773014,
11633007, 11851305, 11503008, 11590781, and 11203013.
P.Z.\ acknowledges support from the NWO Veni Fellowship, grant
NO.\ 639.041.647.
B.J.\ acknowledges Jiangsu NSF grant BK20141310 and
SRFDP of China 20110091120001.
\end{acknowledgements}

\label{lastpage}
\end{document}